\documentstyle[12pt]{article}
\def\NPB#1#2#3{Nucl. Phys. B{#1}, #3 (19#2)}
\def\PLB#1#2#3{Phys. Lett. B{#1}, #3 (19#2)}

\def\PRD#1#2#3{Phys. Rev. D{#1}, #3 (19#2)}
\def\PRL#1#2#3{Phys. Rev. Lett. {#1}, #3 (19#2)}

\begin{document}
\makeatletter \@addtoreset{equation}{section}
\makeatother
\renewcommand{\theequation}{\thesection.\arabic{equation}}
\begin{titlepage}
\begin{flushright}
HUE-99/1 \\
July 1999
\end{flushright}
\bigskip
\bigskip
\begin{center}
{\Large \bf
{\it D} = 4, {\it N} = 1, Type IIA Orientifolds  \\ 
}
\bigskip
S. Ishihara, H. Kataoka and Hikaru Sato\\
\bigskip
{\small \it
Department of Physics,\\
Hyogo University of Education,\\
Yashiro-cho, Hyogo 673-1494, JAPAN
}
\end{center}
\bigskip
\bigskip 
\bigskip
\begin{abstract}
We study $D=4, N=1$, type IIA orientifold with orbifold group $Z_N$ and $Z_N \times Z_M$. We calculate one-loop vacuum amplitudes for Klein bottle, cylinder and M\"obius strip and extract the tadpole divergences. We find that the tadpole cancellation conditions thus obtained are satisfied by the $Z_4$, $Z_8$, $Z_8^\prime$, $Z_{12}^\prime$ orientifolds while there is no solution for $Z_3$, $Z_7$, $Z_6$, $Z_6^\prime$, $Z_{12}$. The $Z_4 \times Z_4$ type IIA orientifold is also constructed by introducing four different configurations of 6-branes. We argue about perturbative versus non-perturbative orientifold vacua under T-duality between the type IIA and the type IIB $Z_N$ orientifolds in four dimensions.
\end{abstract}
\end{titlepage}
%
\section{Introduction}

   In the past years various dualities in ten dimensions among type IIA, type IIB, $E_8 \times E_8$ heterotic, Spin(32)/Z$_2$ heterotic and type I superstring theories have been studied extensively. Most of these dualities are non-perturbative and often allow mapping non-perturbative phenomena in one theory to perturbative phenomena in another theory \cite{pol}.
For example, type I vacua are supposed to be S-dual to strongly coupled $SO(32)$ heterotic vacua and hence there is a hope to get information about non-perturbative heterotic physics through the study on the type I vacua \cite{pw}.
Type I strings can be understood as an orientifold of type IIB closed strings with respect to the world-sheet parity operation $\Omega$.

The type IIA and type IIB theories in $D=10$ are T-dual each other with respect to one compact dimension and these two theories are two limiting points in a continuous moduli space of quantum vacua.  The two heterotic theories are also T-dual each other, though there are technical details involving Wilson loops \cite{ginsparg}.  T-duality applied to the type I theory gives a dual description, which is called type I$^{\prime}$. The type I$^{\prime}$ theory is an orientifold quotient of type IIA theory with respect to the composition of world-sheet parity $\Omega$ with the space reflection $R$.

String vacua of type IIB orientifolds in less than ten dimensions, especially in $D = 6$ \cite{D=6_I,gp,D=6_IIB} and $D = 4$ \cite{D=4,ald,kak} have been studied by many authors. It was found that $D = 4, N = 1$ type IIB orientifolds have consistent string vacua for the orientifold group $Z_3,Z_7,Z_6,Z_6^\prime,Z_{12}$, but they are not consistent for  $Z_4,Z_8,Z_8^\prime,Z_{12}^\prime$. In the heterotic orbifolds, all these $Z_N$ actions can give rise to consistent string vacua \cite{dhvw,imnq}
so that it seems puzzle that some of $Z_N$ are not allowed in the type IIB orientifolds. This puzzle was studied by Ref.\cite{kst}. According to their analysis, additional non-perturbative sectors appear in the four-dimensional type IIB $Z_4,Z_8,Z_8^\prime,Z_{12}^\prime$ orientifolds and the naive tadpole cancellation conditions are not necessarily satisfied. On the other hand, the type IIB $Z_3,Z_7,Z_6,Z_6^\prime,Z_{12}$ orientifolds can obey the perturbative tadpole cancellation conditions although there are some subtleties in the $Z'_6$ case \cite{kst}. 

Despite the extensive studies on the type IIB orientifolds, $D = 4, N = 1$ type IIA orientifold models or the type I$'$ theory in four dimensions have not been constructed explicitly. In this paper we undertake a systematic study of this class of orientifolds.
We present a detailed study of tadpole cancellation conditions for general $D=4$, type IIA orientifolds and explicitly construct the massless spectrum of all possible $Z_N$ orientifolds with 6 branes sitting at the fixed point at the origin. We find that the tadpole cancellation conditions allow the $Z_4$, $Z_8$, $Z_8^\prime$, $Z_{12}^\prime$ orientifolds, while the $Z_3$, $Z_7$, $Z_6$, $Z_6^\prime$, $Z_{12}$ are not allowed. Extension of our argument to $Z_N \times Z_M$ orientifolds is straightforward. As an example, we give the $Z_4 \times Z_4$ orientifold model.  

For the D6-brane configuration of the type IIA orientifolds, each of the three complex planes of the 6-D compact space has two target space coordinates where corresponding open string wave functions $X^{\mu}(\tau,\sigma)$ have the Neumann and the Dirichlet boundary conditions, respectively.
These string wave functions with the different boundary conditions are mixed under the $Z_N$ orbifold action. 
Then, $Z_N$ invariance requires that we should include the open string sectors with the mixed boundary conditions which are neither Neumann nor Dirichlet. These open strings do not end on D-branes and are not dealt with by a world-sheet, {\it i.e.}, perturbative description within the orientifold approach \cite{kst}. It is shown that for the $Z_4,Z_8,Z_8^\prime,Z_{12}^\prime$ orientifolds, we can eliminate mixed boundary conditions by choosing appropriate coordinate axes of the 6-D compact space. Thus, for these orientifolds, there are no non-perturbative states in the above sense and they obey the perturbative tadpole cancellation conditions. On the other hand, for the $Z_3,Z_7,Z_6,Z_6^\prime,Z_{12}$ orientifolds it is shown that we cannot eliminate mixed boundary conditions. Therefore, these orientifolds may have additional non-perturbative states from the open strings with mixed boundary conditions and do not satisfy perturbative tadpole cancellation conditions.

This result is opposite of what was obtained in the $D=4$ type IIB orientifolds: The four dimensional type IIB orientifold vacua have a perturbative solution for $Z_3$, $Z_7$, $Z_6$, $Z_6^\prime$, $Z_{12}$ and non-perturbative for $Z_4$, $Z_8$, $Z_8^\prime$, $Z_{12}^\prime$ \cite{kst}. On the other hand, type IIA orientifold vacua have a perturbative solution for $Z_4$, $Z_8$, $Z_8^\prime$, $Z_{12}^\prime$, while non-perturbative for $Z_3$, $Z_7$, $Z_6$, $Z_6^\prime$, $Z_{12}$.
We expect that $D = 4, N = 1$ type IIA orientifold is T-dual to $D = 4, N = 1$ type IIB orientifold with respect to $X_5, X_7, X_9$ directions of the 6-D orbifold space. What we find is that the perturbative (non-perturbative) vacua of the $D=4$ type IIA orientifolds are changed to the non-perturbative (perturbative) vacua of the $D=4$ type IIB orientifolds under the T-dual transformation.

The remainder of this paper is organized as follows. In Section 2 we summarize how to construct type IIA orientifolds in four dimensions. This is a straightforward extension of the method to construct type IIB orientifolds. In Section 3 we calculate the one-loop vacuum amplitudes and extract the tadpole divergences. The tadpole cancellation conditions are examined in Section 4 for possible $Z_N$ which acts crystallographically on a $T^6$ lattice and lead to $N=1$ supersymmetry. As advertised we find the perturbative solution for  $Z_4,Z_8,Z_8^\prime,Z_{12}^\prime$, while the naive tadpole cancellation conditions are not satisfied for  $Z_3,Z_7,Z_6,Z_6^\prime,Z_{12}$. In Section 5 we discuss our result in the light of T-dualities. We argue that the four-dimensional type IIA orientifolds for $Z_3,Z_7,Z_6,Z_6^\prime,Z_{12}$ have open strings with the mixed boundary conditions which produce the non-perturbative sector. This could be the reason why the naive tadpole cancellation conditions are not satisfied by the $Z_3,Z_7,Z_6,Z_6^\prime,Z_{12}$ models.
Finally, in Appendix we give the details of the calculation for one-loop vacuum amplitudes of the $D=4$ type IIA orientifolds for Klein bottle, cylinder and M\"obius strip surfaces.
%
\section{{\it D}=4, {\it N}=1, Type IIA Orientifolds}

   In this section we summarize the basic ingredients and notation needed in the construction of $D = 4, N = 1$ type IIA orientifolds. The argument is parallel with the type IIB orientifold \cite{gp,ald}. In the type IIA string theory the complete orientifold group is written as $G_1+\Omega RG_2$ with $\Omega R g \Omega R g' \in G_1$ for $g, g' \in G_2$ \cite{gp}.  $\Omega$ is the world-sheet parity transformation that exchanges left and right world sheet movers. $R$ is exchange $X_5 \to -X_5,X_7 \to -X_7,X_9 \to -X_9$, and we are concerned with $G_1=G_2$ and $G_1=Z_N$  or $G_1 = Z_N \times Z_M$ action on $T^6$ in the type IIA string theory.
   
The $Z_N$ orbifold action is realized by powers of the twist generator $\theta (\theta^N=1)$ which is written in the form
\begin{equation}
\theta={\rm exp}(2i\pi(v_1J_{45}+v_2J_{67}+v_3J_{89})),
\label{eq:2.1}
\end{equation}
where $J_{mn}$ are $SO(6)$ Cartan generators. A twist vector $v = (v_1,v_2,v_3)$ associated with $\theta$ obeys $v_1 \pm v_2 \pm v_3 =0$ for some choice of signs in order to realize $N=1$ supersymmetry \cite{dhvw}.
In terms of the complex bosonic coordinates $Y_1=X_4+iX_5,Y_2=X_6+iX_7,Y_3=X_8+iX_9$ and $\bar{Y}_1=X_4-iX_5$ etc..
$\theta$ acts diagonally as
\begin{equation}
\theta^k Y_i=e^{2i\pi kv_i} Y_i, 
\label{eq:2.2}
\end{equation}
and $R$ acts as $RY_i=\bar{Y}_i$.
Similarly, we define complex fermionic fields $\psi^1=\psi^4+i\psi^5$, etc..

To derive the massless spectra we will work in the light-cone gauge.
The Neveu-Schwarz (NS) massless states are  
$\psi_{-\frac{1}{2}}^\mu|0\rangle$  
and $\psi_{-\frac{1}{2}}^i|0\rangle$ which transform as
\begin{eqnarray}
\theta^k \psi_{-\frac{1}{2}}^\mu|0\rangle
&=& \psi_{-\frac{1}{2}}^\mu|0\rangle,
\label{eq:2.3} \\
\theta^k \psi_{-\frac{1}{2}}^i|0\rangle
&=& e^{2i \pi kv_i }\psi_{-\frac{1}{2}}^i|0\rangle,
\label{eq:2.4}
\end{eqnarray}
where $\mu$ denotes uncompactified dimensions.
The massless Ramond (R) states are of the form $|s_0s_1s_2s_3\rangle$ 
with $s_0,s_i=\pm \frac{1}{2}$ and the GSO projection leads to an odd number of minus signs in the right-moving states and an even number of minus signs in the left-moving states or vice versa.
These transform as
\begin{equation}
\theta^k|s_0s_1s_2s_3\rangle = e^{2i\pi kv\cdot s}|s_0s_1s_2s_3\rangle.
\label{eq:2.5}
\end{equation}
The $Z_N$ actions that can act crystallographically on a $T^6$ lattice and lead to $N=1$ supersymmetry were classified in \cite{dhvw}. The list of possible $Z_N$ with corresponding twist vectors is given in Table 1.
%
%
\begin{table}
\begin{center}
\begin{tabular}{|c|c|c|} \hline
\multicolumn{1}{|c|}
{$Z_3:\frac{1}{3}(1,1,-2)$}
&$Z_4:\frac{1}{4}(1,1,-2)$&
$Z_6:\frac{1}{6}(1,1,-2)$
\\ 
\it $Z_6':\frac{1}{6}(1,-3,2)$
&$Z_7:\frac{1}{7}(1,2,-3)$&
$Z_8:\frac{1}{8}(1,3,-4)$
\\ 
\it $Z_8':\frac{1}{8}(1,-3,2)$&
$Z_{12}:\frac{1}{12}(1,-5,4)$&
$Z_{12}':\frac{1}{12}(1,5,-6)$
\\ \hline
\end{tabular}
\end{center}
\caption{$Z_N$ actions in $D=4$}
\end{table}

Although type IIA is a theory of closed strings, the orientifold projection requires both closed and open string sectors.
Tadpole divergences are found in the partition function of the closed sector Klein bottle surface.
To cancel these tadpoles, new contribution must be included \cite{cp}.
Introduction of open strings leads to the required cancellation for a specific structure of Chan-Paton charges.
Tadpole cancellation is achieved by including the right number of D$p$-branes \cite{pol2}.
An open string has one end, labeled by $a$, on a D$p$-brane and the other end, labeled by $b$, on a D$q$-brane.
They give rise to $pq$ string sectors.
The labels $a,b$ correspond to the Chan-Paton factors at each end of the string.

The spectrum in the closed sector is obtained from those of the type IIA orbifold states invariant under $\Omega R$ transformation.
Orbifold states are constructed by coupling left and right moving states of opposite chirality to be invariant under the orbifold group action.
The massless left NS states correspond to vector matters
$\psi_{-\frac{1}{2}}^\mu|0\rangle$ and to scalar matters
$\psi_{-\frac{1}{2}}^i|0\rangle$. Vectors are invariant under the orbifold twist $\theta$ action while scalars acquire a phase $e^{2i \pi kv_i}$. Right movers are obtained by replacing $\psi \to \tilde{\psi}$.

The type of D-branes present in the open string sector depends on the content of the orientifold group. For $G_1 = G_2$, the identity is in $G_2$ so that the orientifold group contains $\Omega R$ as an element.  Since $R$ is an order two element acting on $X_5, X_7, X_9$, there will be D6-branes which extend in $D=4$ space-time plus $X_4,X_6,X_8$ directions and have the Dirichlet(D) condition in $X_5,X_7,X_9$. Furthermore, when $G_1 = G_2 = Z_N$ with $N = $ even, the $Z_N$ action in Table 1 contains order two element $R_3$ acting on the two complex directions transverse to $Y_3$. Then the orientifold group contains an action of the type $\Omega RR_3$ and there is another D6-branes denoted by D$6_3$ (or D$6'$)-branes, which extend in $D=4$ space-time plus $X_5,X_7,X_8$ directions and have the D condition in $X_4,X_6,X_9$.  
In what follows we consider mainly $6_3$-branes so that we denote it as $6'$-branes.
If we consider $Z_N \times Z_M$ with twist vectors $v_\theta = \frac{1}{N}(1,-1,0)$ and $v_\omega = \frac{1}{M}(0,1,-1)$, there is a set of $6_1, 6_2, 6_3$-branes where $6_1$-branes have the D condition in $X_5,X_6,X_8$ directions and $6_2$-branes have the D condition for $X_4,X_7,X_8$.

Open string states are denoted by $|\Psi,ab\rangle$, where $\Psi$ refers to world-sheet degrees of freedom while the Chan-Paton indices $a,b$ are associated to the string endpoints on D$p$-branes and D$q$-branes.
A pair of Chan-Paton labels must be contracted with a hermitian matrix $\lambda_{ab}$. 

The action of a group element $g \in G_1$ is given by
\begin{equation}
g:|\Psi,ab\rangle \to (\gamma_{g,p})_{aa'}|g\Psi,a'b'\rangle(\gamma_{g,q}^{-1})_{b'b},
\label{eq:2.6}
\end{equation}
where $\gamma_{g,p}$ and $\gamma_{g,q}$ are unitary matrices associated to $g$.
The action of $\Omega Rg$ with $g \in G_2$, is given by

\begin{equation}
\Omega Rg:|\Psi,ab\rangle \to
(\gamma_{\Omega Rg,p})_{aa'}|g\Psi,b'a'\rangle(\gamma_{\Omega R g,q}^{-1})_{b'b}.
\label{eq:2.7}
\end{equation}
$\gamma_{\Omega Rg,p}$ is defined as
\begin{equation}
\gamma_{\Omega Rg,p}=\gamma_{g,p}\gamma_{\Omega R,p}=\gamma_{g,p}\gamma_{R,p}\gamma_{\Omega,p}.
\label{eq:2.8}
\end{equation}
Matrices $\gamma_{\Omega R,p}$ and $\gamma_{\Omega Rg,p}$ are unitary.
For $\theta^k \in Z_N$ we abbreviate $\gamma_{\theta^k,p}$ as $\gamma_{k,p}$ and $\gamma_{\Omega R\theta^k,p}$ as $\gamma_{\Omega Rk,p}$.
Since $\Omega R \theta^k (\Omega R)^{-1} = \theta^{N-k}$ so that $(\Omega R \theta^k)^2=1$, Eq.(\ref{eq:2.7}) leads
\begin{equation}
\gamma_{\Omega Rk,p}^{-1}\gamma_{\Omega Rk,p}^T=\pm 1,
\end{equation}
\label{eq:2.9}
then
\begin{equation}
\gamma_{\Omega Rk,p}^T=\pm \gamma_{\Omega Rk,p}.
\label{eq:2.10}
\end{equation}
Consistency with the orientifold group multiplication law implies several constraints on the $\gamma$ matrices.
Without loss of generality we can choose $\gamma_{0,p}=1$ and  
\begin{equation}
 \gamma_{k,p}=\gamma_{1,p}^k, ~~~~\gamma_{1,p}^N = \pm 1.
\label{eq:2.11} 
\end{equation}
Cancellation of tadpoles imposes further conditions on the $\gamma$ matrices. For example, as will be shown in Sec.4, it turns out that the $\gamma$ matrices are $32 \times 32$ and obey the following relations \cite{gp},
\begin{equation}
\gamma_{\Omega R,6}^T=\gamma_{\Omega R,6}, ~~~~\gamma_{\Omega R,6'}^T= -\gamma_{\Omega R,6'}.
\label{eq:2.12}
\end{equation}

The open string spectrum is computed once the $\gamma$ matrices are found.
According to the end point there are various $pq$ sectors.
Here we will concentrate on models containing 6-and $6'$-branes, located on the fixed point corresponding to the origin in the compact space. In this configuration one gets maximal gauge symmetry.
We describe the massless bosonic states in each $pq$ sector.
For 66- or $6'6'$-states, massless NS states include gauge bosons $\psi_{-\frac{1}{2}}^\mu|0,ab\rangle \lambda_{ab}^{(0)}$ 
and scalar matters
$\psi_{-\frac{1}{2}}^i|0,ab\rangle \lambda_{ab}^{(i)}$.
The invariance under the orientifold group action (\ref{eq:2.6}) and (\ref{eq:2.7}) leads to the constraints on the Chan-Paton matrices as
\begin{equation}
\begin{array}{lcl}
 \lambda^{(0)} = \gamma_{1,6} \lambda^{(0)} \gamma_{1,6}^{-1} &,
 &\lambda^{(0)} = -\gamma_{\Omega R,6} \lambda^{(0)^{T}} \gamma_{\Omega R,6}^{-1} \\
 \lambda^{(i)} = e^{2\pi iv_i}\gamma_{1,6} \lambda^{(i)} \gamma_{1,6}^{-1} &,
&\lambda^{(i)} = -\gamma_{\Omega R,6} \lambda^{(i)^T} \gamma_{\Omega R,6}^{-1} 
\end{array}
\label{eq:2.13}
\end{equation}
for 66-states.  
As for $6'6'$-states, we have
\begin{equation}
\begin{array}{lcl}
 \lambda^{(0)} = \gamma_{1,6'} \lambda^{(0)} \gamma_{1,6'}^{-1} &,
 &\lambda^{(0)} =-\gamma_{\Omega R,6'} \lambda^{(0)^{T}} \gamma_{\Omega R,6'}^{-1} \\
\lambda^{(3)} = e^{2\pi iv_i}\gamma_{1,6'} \lambda^{(3)}\gamma_{1,6'}^{-1} &,
&\lambda^{(3)} = -\gamma_{\Omega R,6'} \lambda^{(3)^T} \gamma_{\Omega R,6'}^{-1} 
\end{array}
\label{eq:2.14}
\end{equation}
and for $j=1,2$
\begin{equation}
\lambda^{(j)} = e^{2\pi iv_j}\gamma_{1,6'} \lambda^{(j)} \gamma_{1,6'}^{-1} ~~, 
~~\lambda^{(j)} = \gamma_{\Omega R,6'} \lambda^{(j)^T} \gamma_{\Omega R,6'}^{-1}.
\label{eq:2.15}
\end{equation}

In Eq.(\ref{eq:2.13})-Eq.(\ref{eq:2.15}), the sign in the $\Omega R$ projection is determined as follows \cite{gp}: For the 66- or $6'6'$-sector, the massless states with vertex operator $\partial_t X^\mu$ for $\mu$ parallel to the brane has $\Omega = -1$, and the states with $\partial_n X^\mu$ for $\mu$ perpendicular has $\Omega = +1$. Thus $\Omega R = -1$ for $\mu = 2,3$ and $\mu = 4,5,6,7,8,9$, or $i = 1,2,3$ in the 66-states. For the $6'6'$-states, $\Omega R = +1$ for $\mu = 4,5,6,7$, or $i = 1,2$ and $\Omega R = -1$ for $\mu = 8,9$, or $i = 3$.

Next let us consider the $66'$ states. In this case $X_4,X_5,X_6,X_7$ obey DN boundary conditions and have expansions with half-integer moded creation operators.
By world-sheet supersymmetry their fermionic partners in the NS sector are integer moded.
Their zero modes span a representation of a Clifford algebra and are labeled as $|s_j,s_k\rangle, j,k=1,2$ with $s_j,s_k=\pm \frac{1}{2}$.
Under $\theta ,|s_j,s_k\rangle$ picks up a phase $e^{2\pi i(v_js_j+v_ks_k)}$.
Hence, for these states we have
\begin {equation}
\lambda=e^{2\pi i(v_js_j+v_ks_k)}\gamma_{1,6} \lambda \gamma_{1,6'}^{-1}.
\label{eq:2.16}
\end{equation}
The $6'6$ sectors are related to $66'$ by $\Omega$ so that we have no extra constraints on $\lambda$ in the $6'6$ sectors. 

Now let us consider how to extract spectrum of the 66-states. 
Since $\gamma_{\Omega R,6}$ are symmetric from Eq.(\ref{eq:2.12}), the constraints on $\lambda^{(0)}$ and $\lambda^{(i)}$ under $\Omega R$ in Eq.(\ref{eq:2.13}) implies that  $\lambda^{(0)}$ and $\lambda^{(i)}$ are $SO(32)$ generators. They can be organized into charged generators $\lambda_a=E_a,a=1,\cdots ,480,$ and Cartan generators $\lambda_I=H_I,I=1,\cdots,16$ such that 
\begin {equation}
[H_I,E_a]=\rho_I^aE_a,
\label{eq:2.17}
\end{equation}
where $(\rho_1^a,\cdots,\rho_{16}^a)$ is the 16-dimensional root vector associated to the generator $E_a$.
These vectors are of the form $(\underline{\pm 1,\pm 1,0,\cdots,0})$, where underlining indicates that all possible permutations must be considered.

The $\gamma_{1,6}$ and its powers represent the action of $Z_N$ group on Chan-Paton factors, and they correspond to elements of a discrete subgroup of the Abelian group spanned by the Cartan generators.
We can write 
\begin {equation}
\gamma_{1,6}=e^{-2i\pi V_{66}\cdot H}.
\label{eq:2.18}
\end{equation}
This equation defines the $16$-dimensional shift vector $V_{66}$.
Gauge bosons are selected by the first condition of Eq.(\ref{eq:2.13}). They are given by 
\begin {equation}
\rho^a \cdot V_{66}=0  \bmod {\bf Z},
\label{eq:2.19}
\end{equation}
which select the subgroup of $SO(32)$ \cite{ald}. 
Matter states are selected by the equation for $\lambda^{(i)}$ in Eq.(\ref{eq:2.13}) and are given by
\begin{equation}
\rho^a \cdot V_{66}=v_i \bmod {\bf Z}.
\label{eq:2.20}
\end{equation}

For the $6'6'$ sector, $\gamma_{\Omega R,6'}$ is antisymmetric from Eq.(\ref{eq:2.12}). Then $\lambda^{(0)}$ and $\lambda^{(3)}$ are $Sp(32)$ generators which are associated with the same $SO$ root vectors given before plus long roots $(\underline{\pm 2,0,\cdots,0})$. 
On the other hand, $\lambda^{(j)}, j=1,2$ are $SO(32)$ generators due to Eq.(\ref{eq:2.15}).
The $\gamma_{1,6'}$ is expressed as Eq.(\ref{eq:2.18}) with the shift vector $V_{6'6'}$. Then the gauge symmetry and spectra of matter states are determined by the same form of Eqs.(\ref{eq:2.19}) and (\ref{eq:2.20}) for $V_{6'6'}$, respectively. 

For the $66'$ sector, we have generators acting simultaneously on both 6-branes and $6'$-branes. For the ground states $|s_j,s_k \rangle$, the roots of the generators are given by
\begin {equation}
\rho_{66'}=(\underline{\pm1,0,\cdots,0}
;\underline{\pm 1,0,\cdots,0}),
\label{eq:2.21}
\end{equation}
where the first 16 components transform under $SO(32)$ of 6-branes and second 16 components transform under $SO(32)$ of $6'$-branes. 
The shift in this sector is defined to be 
$V_{66'}=V_{66} \otimes V_{6'6'}$.
Massless states of the 6$6'$ sector are determined by Eq.(\ref{eq:2.16}) and are given by
\begin {equation}
\rho_{66'}\cdot V_{66'}=(s_jv_j+s_kv_k)\bmod {\bf Z},
\label{eq:2.22}
\end{equation}
with $s_j, s_k=\pm \frac{1}{2}$, where the GSO projection imposes $s_j = s_k$ and minus sign corresponds to antiparticles.
%
\section{Tadpoles}

In the orientifold theory the one-loop vacuum amplitudes include the torus, the Klein bottle$({\cal K})$, the M\"{o}bius strip$({\cal M})$, and the cylinder$({\cal C})$.
The last three have tadpole divergences from exchange of massless states in the closed string channels.
By supersymmetry the total divergences vanish but consistency requires separate cancellation of NS-NS and R-R tadpoles \cite{cp}.
For $D=4,N=1$,type IIB orientifold, tadpole cancellation conditions have been studied in Refs.\cite{D=4,ald}. In this section we summarize the result of the calculation for the tadpoles of the type IIA $Z_N$ orientifolds. Details are given in the appendix.

(1) Klein bottle amplitude

The Klein bottle amplitude is given by
\begin {equation}
{\cal K}=\frac{V_4}{8N}\sum_{n,k=0}^{N-1}
\int_0^\infty\frac{dt}{t}Z_{\cal K} (\theta^n,\theta^k),
\label{eq:3.1}
\end{equation}
where
\begin {equation}
Z_{\cal K} (\theta^n,\theta^k) = {\rm Tr}\left[\frac{1+(-1)^F}{2}\frac{1\pm(-1)^{\tilde{F}}}{2}\Omega R \theta^k e^{-2\pi t[L_0(\theta^n)+{\tilde L}_0(\theta^n)]}\right].
\label{eq:3.2}
\end{equation}
Here $L_0(\theta^n)$ and $\tilde{L}_0(\theta^n)$ are the Virasoro operators from left-moving and right-moving modes in the twisted sector $n$, respectively. $F (\tilde{F})$ is the left (right)-moving world-sheet fermion number. Since in the type IIA theory Ramond states have opposite chirality in left- and right-moving sectors, the sign in front of $(-1)^{\tilde{F}}$ is $+$ for the NS sector and $-$ for the Ramond sector. $V_4$ denotes the regularized 4-D space-time volume.

The trace in $Z_{\cal K}$ is computed in a standard way. Since $\Omega R$ exchanges the $\theta^n$-twisted sector with itself, i.e.,
\begin{equation}
\Omega R L_0(\theta^n) (\Omega R)^{-1} = L_0(\theta^{n})
\label{eq:3.3}
\end{equation}
and similarly for $\tilde{L}_0(\theta^n)$, the $\theta^n$-twisted sector survives the trace, in general. The divergences of interest are produced by the $t \rightarrow 0$ limit. For the untwisted ($n = 0$) sector, we obtain
\begin{equation}
Z_{\cal K}(1,\theta^k) \rightarrow  (1 - 1) \frac{32}{\pi^4}t^2 \prod_i \frac{1}{t}\frac{L_{e_i}}{L_{o_i}} \prod_j\frac{1}{t}\frac{L_{o_j}}{L_{e_j}},
\label{eq:3.4}
\end{equation}
where $i, j$ denote the complex planes with $kv_i$ = integer and $kv_j$ = half-integer. The length of the compact space is denoted by $ L_{e_i} = (L_4,L_6,L_8)$ and $ L_{o_i} = (L_5,L_7,L_9)$.

For the twisted sector, it turns out that 
\begin {equation}
Z_{\cal K}(\theta^{n},\theta ^k) = -Z_{\cal K}(\theta^{N-n},\theta ^k),
\label{eq:3.5}
\end{equation}
so that the twisted sectors give no contribution to the Klein bottle amplitude (\ref{eq:3.1}).

(2) Cylinder amplitude

The cylinder amplitude is given by
\begin {eqnarray} 
{\cal C}_{pq}&=&\frac{V_4}{8N}\sum_{k=0}^{N-1}
\int_0^\infty\frac{dt}{t}Z_{pq}(\theta^k),
\label{eq:3.6}\\
Z_{pq}(\theta^k)&=&{\rm Tr}\left[\frac{1+(-1)^F}{2}\theta^k e^{-2\pi t L_0}\right].
\label{eq:3.7}
\end{eqnarray}
The trace is over open string states with boundary conditions according to the D$p$ and D$q$-branes at the endpoint. There are three types of $Z_{pq}$ ; $Z_{66}, Z_{6'6'}, Z_{66'}$. 

For $ Z_{66}$ in the limit $t \rightarrow 0$, we obtain
\begin{equation}
Z_{66}(\theta^k) \rightarrow (1 - 1) \frac{1}{16\pi^4 t} \prod_i \frac{L_{e_i}}{L_{o_i}}\prod_j 2|\sin\pi kv_j|  \sum_I ({\rm Tr}\gamma_{k,6,I})^2,
\label{eq:3.8}
\end{equation}
where $i$ and $j$ are for the planes with $kv_i =$ integer and $kv_j \neq$  integer, respectively. $I$ refers to the fixed points of $\theta^k$.

In a similar way, the limit of $Z_{6'6'}$ is obtained as
\begin{equation}
Z_{6'6'}(\theta^k) \rightarrow (1 - 1) \frac{1}{16\pi^4 t} \prod_i \frac{L_{m_i}}{L_{\ell_i}}\prod_j (-1)^{p_j}2\sin\pi kv_j \sum_I ({\rm Tr}\gamma_{k,6',I})^2,
\label{eq:3.9}
\end{equation}
where $L_{m_i} = (L_5, L_7, L_8)$ and $L_{\ell_i} = (L_4, L_6, L_9)$.  This is due to the fact that the $6'6'$-states have NN boundary conditions for $\mu = 5,7,8$ directions and DD boundary conditions for $\mu = 4,6,9$ directions. In the product with respect to $j$, $p_j$ is given such that $p_j < kv_j < p_j +1, p_j =$ integer. Other notations are the same as $Z_{66}$.

Next, let us consider $Z_{66'}$.  The $66'$-states have the DN boundary conditions for $\mu = 4,5,6,7$ directions and the NN boundary condition  @for $\mu = 8$ and the DD boundary condition for $\mu = 9$. The $t \rightarrow 0$ limit of $Z_{66'}$ vanishes for $kv_i =$ integer, $i = 1,2$. When $kv_i \neq$ integer, $i = 1,2,3$, it is given by
\begin{equation}
Z_{66'}(\theta^k) \rightarrow (1 - 1) \frac{1}{16\pi^4 t} 2|\sin\pi kv_3| \prod_{i=1,2} (-1)^{p_i +1}\sum_I ({\rm Tr}\gamma_{k,6,I})\sum_{I'} ({\rm Tr}\gamma_{k,6',I'})
\label{eq:3.10}
\end{equation}
for $p_j < kv_j < p_j +1, p_j =$ integer. When $kv_i \neq$ integer, $i = 1,2$ and $kv_3 =$ integer, the limit of $Z_{66'}$ reads
\begin{equation}
Z_{66'}(\theta^k) \rightarrow (1 - 1) \frac{1}{16\pi^4 t} \frac{L_8}{L_9} \prod_{i=1,2} (-1)^{p_i +1}\sum_I ({\rm Tr}\gamma_{k,6,I})\sum_{I'} ({\rm Tr}\gamma_{k,6',I'}).
\label{eq:3.11}
\end{equation}
(3) M\"obius strip amplitude

M\"{o}bius strip amplitude is given by
\begin {eqnarray}
{\cal M}_p=\frac{V_4}{8N}\sum_{k=0}^{N-1}
\int_0^\infty\frac{dt}{t}Z_p(\theta^k),
\label{eq:3.12}\\
Z_p(\theta^k)={\rm Tr}[\frac{1+(-1)^F}{2}\Omega R\theta^k e^{-2\pi tL_0}],
\label{eq:3.13}
\end{eqnarray}
where $L_0$ has the same form as the cylinder amplitude.

M\"obius 6-states have NN boundary conditions for $\mu$ = 4,6,8 directions and DD boundary conditions for $\mu =$ 5,7,9. The $t \rightarrow 0$ limit of $Z_6(\theta^k)$ amounts to
\begin{equation}
Z_6(\theta^k) \rightarrow (1-1)\frac{-1}{8\pi^4 t}\prod_i (-1)^{kv_i}\frac{2L_{e_i}}{L_{o_i}}\prod_j (-1)^{p_j} 2\sin \pi kv_j\sum_I {\rm Tr}(\gamma^{-1}_{\Omega Rk,6,I}\gamma^{T}_{\Omega Rk,6,I})
\label{eq:3.14}
\end{equation}
where $i$ is for the plane with $kv_i =$ integer and $j$ is for the plane with $2kv_j \neq$ integer,  $p_j < 2kv_j < p_j +1$. When there is a plane with $kv_i =$ half integer, the limit of $Z_6(\theta^k)$ vanishes.

M\"obius $6'$-states have NN boundary conditions for $\mu =$ 5,7,8 directions and DD boundary conditions for $\mu =$ 4,6,9. When $2kv_j \neq$ integer, $j = 1,2,3$, the $t \rightarrow 0$ limit of $Z_{6'}(\theta^k)$ is given by
\begin{eqnarray}
Z_{6'}(\theta^k) &\rightarrow& (1-1)\frac{-1}{8\pi^4 t}(-1)^{p_3}2\sin\pi kv_3 \prod_{i=1,2} (-1)^{p_i +1}2\cos\pi kv_i 
\nonumber\\
&& \times \sum_I {\rm Tr}(\gamma^{-1}_{\Omega Rk,6',I}\gamma^{T}_{\Omega Rk,6',I}),
\label{eq:3.15}
\end{eqnarray}
where $p_j < 2kv_j < p_j +1, j=1,2,3$. When the following condition for $kv_j$ is satisfied, the corresponding part of Eq.(\ref{eq:3.15}) should be changed as
\begin{equation}
\begin{array}{lclcl}
kv_3 = {\rm integer} &:& (-1)^{p_3}2\sin\pi kv_3 & \rightarrow & (-1)^{kv_3}\frac{2L_8}{L_9} \\
2kv_i = 4n+1 &:& (-1)^{p_i +1}2\cos\pi kv_i & \rightarrow & - \frac{2L_{o_i}}{L_{e_i}} \\
2kv_i = 4n+3 &:& (-1)^{p_i +1}2\cos\pi kv_i & \rightarrow &  \frac{2L_{o_i}}{L_{e_i}}
\end{array}
\label{eq:3.16}
\end{equation}
where $n = $ integer and $L_{o_i} = (L_5, L_7), L_{e_i} = (L_4, L_6)$. For other $kv_j$, i.e., $kv_3 =$ half integer and/or $kv_{1,2} =$ integer, we have $Z_{6'}(\theta^k) \rightarrow (1-1)0$.
%
%
\section{Models}

In this section we study the type IIA orientifolds based on $T^6/\{\Omega R,G\}$ where $G$ denotes generators of a discrete group. Let us consider $G = Z_N$ since generalization to $G = Z_N \times Z_M$ is straightforward. There are 6-branes for all $N$ and in addition there are $6'$-branes for even $N$.
Here we concentrate on models with all the branes located on the particular fixed point corresponding to the origin in the compact space.
In this configuration one gets maximal gauge symmetry.

The various tadpole divergences can be classified according to their volume dependence.  To extract the divergences in the various type of amplitudes, we have to make the change of variables, since the loop modulus $t$ is related to the cylinder length $\ell$ differently for each surface as follows:
\begin{equation}
{\rm Klein \:bottle}: t = \frac{1}{4\ell}, ~~{\rm Cylinder}: t = \frac{1}{2\ell}, ~~{\rm M\ddot{o}bius}: t = \frac{1}{8\ell}.
\label{eq:4.1}
\end{equation}

First, let us consider $Z_N$ with $N$ = odd. In the ${\cal K}$ amplitude, Eq.(\ref{eq:3.4}) shows that $Z_{\cal K} (1,1)$ has tadpole singularity proportional to $L_4L_6L_8/L_5L_7L_9$.
Taking into account the $Z_{66}(1)$ in ${\cal C}_{66}$ and the $Z_6(1)$ in ${\cal M}_6$ which are given by Eq.(\ref{eq:3.8}) and Eq.(\ref{eq:3.14}), respectively, the total amplitude for large $\ell$ is given by
\begin{eqnarray}
{\cal K}+{\cal C}_{66}+{\cal M}_6 &=& (1-1)\int_0^\infty d\ell 
\nonumber \\
&\times& 
\frac{V_4}{2\pi^4 N}\left\{32+\frac{1}{32}({\rm Tr}\gamma_{0,6})^2
-2{\rm Tr}(\gamma_{\Omega R,6}^{-1}\gamma_{\Omega R,6}^T)\right\}\frac{ L_4L_6L_8}{ L_5L_7L_9}.
\label{eq:4.2}
\end{eqnarray}
Since $\gamma_{0,6} = 1$, so Tr($\gamma_{0,6}$) = $n_6$ and $n_6$ stands for the number of 6-branes. Then tadpole divergences are canceled if $n_6 = 32$ and 
\begin{equation}
\gamma_{\Omega R,6}=\gamma_{\Omega R,6}^T.
\label{eq:4.3}
\end{equation} 

For $Z_N$ with $N =$ even, the Klein bottle amplitude ${\cal K}$ has additional tadpole divergence which arises from $Z_{\cal K}(1,\frac{N}{2})$. This divergence is proportional to $L_5L_7L_8/L_4L_6L_9$. For $N =$ even, there are $6'$-branes and this type of divergence will be canceled by the divergences which arise from $Z_{6'}(\frac{N}{2})$ in ${\cal M}_{6'}$ and the $Z_{6'6'}(1)$ in ${\cal C}_{6'6'}$.  In fact, for the twist vectors given in Table 1, the divergent part of the total amplitude is proportional to

\begin {equation}
\left\{32+\frac{1}{32}({\rm Tr}\gamma_{0,6'})^2
+2{\rm Tr}(\gamma_{\Omega R\frac{N}{2},6'}^{-1}\gamma_{\Omega R \frac{N}{2},6'}^T)\right\}\frac{ L_5L_7L_8}{ L_4L_6L_9}
\label{eq:4.4}
\end{equation} 
and vanishes provided that Tr($\gamma_{0,6'}) = n_{6'} = 32$ and 
\begin{equation}
\gamma_{\Omega R\frac{N}{2},6'}=-\gamma_{\Omega R\frac{N}{2},6'}^T.
\label{eq:4.5}
\end{equation}

The action of $\Omega R$ on the 66, $6'6'$ and $66'$ sectors can be analyzed in much the same way as in the type IIB theory for $\Omega$. Following the analysis given by Gimon and Polchinski (GP) \cite{gp}, we can show that $(\Omega R)^2 = 1$ on 66 and $6'6'$ states, whereas $(\Omega R)^2 = -1$ on $66'$ states. Then Eq.(\ref{eq:4.3}) implies
\begin{equation}
\gamma_{\Omega R,6'}= -\gamma_{\Omega R,6'}^T.
\label{eq:4.6}
\end{equation} 

Since the matrix $\gamma_{1,p}$ and its powers $\gamma_{k,p}$ represent the action of the $Z_N$ group on Chan-Paton factors, they can be expressed by the diagonal matrices. Taking into account Eqs.(\ref{eq:2.8}), (\ref{eq:2.10}) and Eqs.(\ref{eq:4.3}), (\ref{eq:4.6}), we can derive the following relations,
\begin{eqnarray}
\gamma^T_{\Omega Rk,p} = \epsilon_p \gamma_{\Omega Rk,p} &{\rm if}& [\gamma_{1,p}, \gamma_{\Omega R,p}] = 0,
\label{eq:4.7}\\
\gamma^T_{\Omega Rk,p} = (-1)^k \epsilon_p \gamma_{\Omega Rk,p} &{\rm if}& \{\gamma_{1,p}, \gamma_{\Omega R,p}\} = 0,
\label{eq:4.8}
\end{eqnarray}
where $\epsilon_p = 1$ for $p = 6$ and $\epsilon_p = -1$ for $p = 6'$. Equation (\ref{eq:4.8}) holds for even $N$. 

Now that we have prepared necessary ingredients to study the $Z_N$ models, we now consider each orientifold in detail.

(1) $Z_3$ and $Z_7$

Possible $Z_N$ models with $N =$ odd are $Z_3$ and $Z_7$.  We consider here the models in which all 6-branes sit at the origin in the compact space. As expressed in Eq.(\ref{eq:4.2}), tadpole divergences in the Klein bottle amplitude ${\cal K}$ are canceled by the divergences in the cylinder amplitude ${\cal C}_{66}$ and the M\"obius strip amplitude ${\cal M}_{6}$ with introducing 32 6-branes. 

There remain, however, extra divergences in the cylinder amplitude ${\cal Z}_{66}(\theta^k)$ and the M\"obius strip amplitude ${\cal Z}_6(\theta^k)$, where $k =1,2$ for $Z_3$ and $k=1,\cdots,6$ for $Z_7$. Using Eq.(\ref{eq:3.8}) and Eq.(\ref{eq:3.14}), we find that they contribute to the coefficient of the divergent part as follows:
\begin{equation}
\sum_{k=1,2}\left[({\rm Tr}\gamma_{k,6})^2 + 8{\rm Tr}(\gamma^{-1}_{\Omega Rk,6}\gamma^T_{\Omega Rk,6})\right],
\label{eq:4.9}
\end{equation}
for $Z_3$ and
\begin{equation}
\sum_{k=1,\cdots,6}\left[({\rm Tr}\gamma_{k,6})^2 - 8{\rm Tr}(\gamma^{-1}_{\Omega Rk,6}\gamma^T_{\Omega Rk,6})\right],
\label{eq:4.10}
\end{equation}
for $Z_7$. 

Let us first consider $Z_3$. Taking into account Eq.(\ref{eq:4.7}), we find the following condition to cancel the tadpole divergences,
\begin{equation}
\sum_{k=1,2}({\rm Tr}\gamma_{k,6})^2 = - 2 \cdot 16^2. 
\label{eq:4.11}
\end{equation}
Since $\gamma_{k,6} = \gamma_{1,6}^k$ and $\gamma_{1,6}$ is $32 \times 32$ diagonal matrix with $\gamma_{1,6}^3 = \pm 1$, we find that there is no such $\gamma_{k,6}$ which obeys Eq.(\ref{eq:4.11}).

The condition of tadpole cancellation for the $Z_7$ orientifold amounts to
\begin{equation}
\sum_{k=1,\cdots,6}({\rm Tr}\gamma_{k,6})^2 =  6 \cdot 16^2.
\label{eq:4.12}
\end{equation}
In general, the diagonal element $\gamma_{ii}$ of $\gamma_{1,p}$ is ${\rm e}^{2\pi ik/N}$ for $\gamma_{1,p}^N = 1$, and ${\rm e}^{i\pi (2k+1)/N}$ for $\gamma_{1,p}^N = -1$, where $k = 1, \cdots, N$. It turns out that under this restriction there is no $\gamma_{k,6}$ which satisfies Eq.(\ref{eq:4.12}).

(2) $Z_4$

This type of orientifold has 32 6-branes and 32 $6'$-branes. Tadpole divergences proportional to $L_4L_6L_8/L_5L_7L_9$ and $L_5L_7L_8/L_4L_6L_9$ are removed by the conditions given by Eq.(\ref{eq:4.3}) and Eq.(\ref{eq:4.5}). In addition, there are tadpole divergences which arise from ${Z}_{pq}(\theta^2), p,q = 6,6'$ and are proportional to $L_8/L_9$. The condition for the cancellation of the divergent part is given by
\begin {equation}
2({\rm Tr}\gamma_{2,6})^2+2({\rm Tr}\gamma_{2,6'})^2+{\rm Tr}\gamma_{2,6}{\rm Tr}\gamma_{2,6'} = 0.
\label{eq:4.13}
\end{equation} 
Further divergences come from  ${Z}_{pq}(\theta^k)$ where $p,q = 6,6'$ and $k = 1,3$. These contributions are canceled provided that
\begin{equation}
\sum_{k=1,3}\left\{({\rm Tr}\gamma_{k,6})^2 +({\rm Tr}\gamma_{k,6'})^2 + ({\rm Tr}\gamma_{k,6})({\rm Tr}\gamma_{k,6'})\right\} = 0.
\label{eq:4.14}
\end{equation}

Equations (\ref{eq:4.13}) and (\ref{eq:4.14}) are satisfied by the following $\gamma$ matrices:
\begin {equation}
\gamma_{1,6} = \gamma_{1,6'}
= \mbox{diag}(\alpha I_8,-I_4,I_4,\alpha^3 I_8,-I_4,I_4),
\label{eq:4.15}
\end{equation}
where $\alpha=e^{2\pi i/4}$ and $I_m$ stands for the $m$ dimensional unit matrix. In fact, for these $\gamma$ matrices we have
\begin{equation}
{\rm Tr}\gamma_{k,6} = {\rm Tr}\gamma_{k,6'} = 0, ~~k = 1,2,3
\label{eq:4.16}
\end{equation}
so that Eqs.(\ref{eq:4.13}) and (\ref{eq:4.14}) are satisfied trivially.@

The $\gamma_{1,p}$, $p =6,6'$ is expressed as Eq.(\ref{eq:2.18}) in terms of Cartan generators $H_I$. Since Cartan generators are represented by $2 \times 2 ~\sigma_3$ submatrices, Eq.(\ref{eq:4.15}) defines the following 16-dimensional shift vector,
\begin {equation}
V_{66} = V_{6'6'}
= \frac{1}{4}(\underbrace{1\cdots 1}_{8},
\underbrace{2\cdots2}_4,\underbrace{0\cdots0}_4) 
\label{eq:4.17}
\end{equation} 
Gauge symmetry is selected by the root vectors $\rho^a$ which obey the condition (\ref{eq:2.19}). We find a $[U(8)\times SO(8)\times SO(8)]$ group from 6-branes and $[U(8)\times Sp(8)\times Sp(8)]$ from $6'$-branes. 

Charged chiral states from the 66 sector are given by Eq.(\ref{eq:2.20}). We find the following 66 matter states,
\begin {equation}
\begin{array}{rrr}
2\,(8,1,8;1,1,1),& 2\,({\bar 8},8,1;1,1,1),& (1,8,8;1,1,1) \\
(28,1,1;1,1,1),& (\overline{28},1,1;1,1,1) & 
\end{array}
\label{eq:4.18} 
\end{equation}
where the first three entries of the parentheses stand for representations with respect to $U(8)\times SO(8)\times SO(8)$ from 6-branes and the last three are for $U(8)\times Sp(8)\times Sp(8)$ from $6'$-branes. In the same way, we obtain the $6'6'$ matter states as follows.
\begin {equation}
\begin{array}{rrr}
2\,(1,1,1;8,1,8),& 2\,(1,1,1;{\bar 8},8,1),& (1,1,1;1,8,8) \\
(1,1,1;28,1,1),& (1,1,1;\overline{28},1,1)& \\
(1,1,1;8,1,1),& (1,1,1;{\bar 8},1,1)&
\end{array}
\label{eq:4.19} 
\end{equation}
Finally, the 6$6'$ matter states are obtained by the condition (\ref{eq:2.22}) as
\begin {equation}
\begin{array}{rr}
(8,1,1;1,1,8),& (1,1,8;8,1,1) \\
(1,8,1;{\bar 8},1,1),& ({\bar 8},1,1;1,8,1)
\end{array}
\label{eq:4.20} 
\end{equation}

(3) $Z_8$

In this case there are tadpole divergences proportional to $L_8/L_9$ which are from ${Z}_{pq}(\theta^k)$ with $p,q = 6,6'; k=2,4,6$ and ${Z}_p(\theta^k)$ with $p = 6,6';k=2,6$. Cancellation of the divergent part gives the following condition.
\begin {eqnarray}
& &\sum_{p=6,6'}\left\{\sum_{k=2,6}({\rm Tr}\gamma_{k,p})^2 +2({\rm Tr}\gamma_{4,p})^2\right\}  + \sum_{k=2,6}{\rm Tr}\gamma_{k,6}{\rm Tr}\gamma_{k,6'} 
 - {\rm Tr}\gamma_{4,6} {\rm Tr}\gamma_{4,6'}
\nonumber\\ 
& & - 16 \sum_{k=2,6}\left\{{\rm Tr}(\gamma_{\Omega Rk,6}^{-1}\gamma_{\Omega Rk,6}^T)
- {\rm Tr}(\gamma_{\Omega Rk,6'}^{-1}\gamma_{\Omega Rk,6'}^T)\right\}
=0.
\label{eq:4.21}
\end{eqnarray} 
There are also tadpole divergences from ${Z}_{pq}(\theta^k)$ with $p,q = 6,6'; k= 1,3,5,7$. Cancellation of these divergences gives the following condition,
\begin{eqnarray}
\sum_{p=6,6'}\sum_{k=1,3,5,7}({\rm Tr}\gamma_{k,p})^2 &-& \sqrt{2}\sum_{k=1,7}({\rm Tr}\gamma_{k,6})({\rm Tr}\gamma_{k,6'})
\nonumber\\
 &+& \sqrt{2}\sum_{k=3,5}({\rm Tr}\gamma_{k,6})({\rm Tr}\gamma_{k,6'})= 0.
\label{eq:4.22}
\end{eqnarray}

Taking into account Eq.(\ref{eq:4.7}) and Eq.(\ref{eq:4.8}) in Eq.(\ref{eq:4.21}), these two conditions of tadpole cancellation are satisfied by the $\gamma$ matrices obeying
\begin{equation}
{\rm Tr}\gamma_{k,6}=0 \,(k\ne 4),
~~{\rm Tr}\gamma_{4,6}=32,
~~{\rm Tr}\gamma_{k,6'}=0
\label{eq:23}
\end{equation}
or the relations obtained by exchanging $6 \leftrightarrow 6'$. The solution of Eq.(\ref{eq:23}) is given by
\begin {equation}
\begin{array}{lcl}
\gamma_{1,6}
&=& \mbox{diag}(iI_8,-I_4,I_4,,-iI_8,-I_4,I_4), \\
\gamma_{1,6'}
&=& \mbox{diag}(\xi I_4,\xi^2I_4,\xi^3I_4,-I_2,I_2,
\xi^7I_4,\xi^6I_4,\xi^5I_4,-I_2,I_2),
\end{array}
\label{eq:4.24}
\end{equation}
where $\xi=e^{2\pi i/8}$.

The $\gamma$ matrices given by Eq.(\ref{eq:4.24}) define the following 16-dimensional shift vector,
\begin {equation}
\begin{array}{lcl}
V_{66} &=& \frac{1}{8}(\underbrace{2\cdots 2}_{8},
\underbrace{4\cdots4}_{4},\underbrace{0\cdots0}_{4}), \\
V_{6'6'} &=& \frac{1}{8}(\underbrace{1\cdots 1}_{4},
\underbrace{2\cdots2}_{4},\underbrace{3\cdots3}_{4},
\underbrace{4\cdots4}_{2},\underbrace{0\cdots0}_{2}).
\end{array} 
\label{eq:4.25}
\end{equation}
Gauge symmetry is determined by the root vectors $\rho^a$ obeying Eq.(\ref{eq:2.19}) and is given by
\begin{equation}
[U(8)\times  SO(8)^2] \times [U(4)^3\times  Sp(4)^2]
\label{eq:4.26}
\end{equation}
where $[U(8)\times  SO(8)^2]$ is from 6-branes and $[U(4)^3\times  Sp(4)^2]$ is from $6'$-branes.

Charged chiral 66 states are obtained by the condition (\ref{eq:2.20}). The 66 states are given by
\begin {equation}
(1,8,8;1^5), ~~(28,1,1;1^5), ~~(\overline{28},1,1;1^5)
\label{eq:4.27}
\end{equation} 
and the $6'6'$ states amount to
\begin {equation}
\begin{array}{rrr}
(1^3;4,1,1,1,4),& (1^3;4,4,1,1,1),& (1^3;{\bar 4},4,1,1,1)\\
(1^3;1,{\bar 4},4,1,1),& (1^3;1,{\bar 4},{\bar 4},1,1),& (1^3;1,1,{\bar 4},4,1)\\
(1^3;1,1,4,1,4),& (1^3;{\bar 4},1,1,4,1),& (1^3;{\bar 4},1,{\bar 4},1,1)\\
(1^3;1,1,1,4,4),& (1^3;4,1,4,1,1),& (1^3;1,{\bar 6},1,1,1)\\
(1^3;1,6,1,1,1),& (1^3;1,4,1,1,1),& (1^3;1,{\bar 4},1,1,1)
\end{array} 
\label{eq:4.28}
\end{equation}
The $66'$ states are selected by the condition (\ref{eq:2.22}) and are given by
\begin {equation}
\begin{array}{rr}
(8,1,1;1,1,1,1,4),& ({\bar 8},1,1;1,1,1,4,1) \\
(1,8,1;1,4,1,1,1),& (1,1,8;1,{\bar 4},1,1,1)
\end{array} 
\label{eq:4.29}
\end{equation}

(4) $Z'_{8}$

In this orientifold model there are tadpole divergences from ${Z}_{pq}(\theta^4), p,q = 6,6'$, which are proportional to $L_8/L_9$. Cancellation of these divergences requires the following condition,
\begin {equation}
2({\rm Tr}\gamma_{4,6})^2+2({\rm Tr}\gamma_{4,6'})^2
+({\rm Tr}\gamma_{4,6})({\rm Tr}\gamma_{4,6'})=0.  
\label{eq:4.30}
\end{equation} 
Other divergences arise from ${Z}_{pq}(\theta^k), p,q = 6,6'; k = 1,\cdots,7 (k \neq 4)$ and ${Z}_p(\theta^k), p = 6, 6'; k = 1,3,5,7$.  These divergences are canceled provided that
\begin{eqnarray}
&&\sum_{p=6,6'}\left\{\sum_{k=1,3,5,7}({\rm Tr}\gamma_{k,p})^2 + 2\sum_{k=2,6}({\rm Tr}\gamma_{k,p})^2 \right\} - \sqrt{2}\sum_{k=1,7}({\rm Tr}\gamma_{k,6})({\rm Tr}\gamma_{k,6'})
\nonumber\\
&& + \sqrt{2}\sum_{k=3,5}({\rm Tr}\gamma_{k,6})({\rm Tr}\gamma_{k,6'}) - 2\sum_{k=2,6}({\rm Tr}\gamma_{k,6})({\rm Tr}\gamma_{k,6'}) \nonumber\\
&&- 8\sum_{k=1,3,5,7}\left\{({\rm Tr}\gamma^{-1}_{\Omega Rk,6}\gamma^T_{\Omega Rk,6}) - ({\rm Tr}\gamma^{-1}_{\Omega Rk,6'}\gamma^T_{\Omega Rk,6'})\right\} = 0.
\label{eq:4.31}
\end{eqnarray}

For the case of Eq.(\ref{eq:4.7}), ${\rm Tr}\gamma^{-1}_{\Omega Rk,6}\gamma^T_{\Omega Rk,6} = - {\rm Tr}\gamma^{-1}_{\Omega Rk,6'}\gamma^T_{\Omega Rk,6'} = 32$ and the above two conditions are satisfied by the following $\gamma$ matrices:
\begin{equation}
\begin{array}{lcl}
{\rm Tr}\gamma_{k,6} &=& {\rm Tr}\gamma_{k,6'} = 0, \mbox{ }k = 1,3,4,5,7 \\
{\rm Tr}\gamma_{2,6} &=& {\rm Tr}\gamma_{2,6'} = 16\sqrt{2} \\
{\rm Tr}\gamma_{6,6} &=& {\rm Tr}\gamma_{6,6'} = -16\sqrt{2}
\end{array}
\label{eq:4.32}
\end{equation}
So, $\gamma_{1,p}, p=6,6'$ is given by
\begin {equation}
\gamma_{1,6}= \gamma_{1,6'}
= \mbox{diag}(\xi I_8,\xi^7I_8,-\xi^7 I_8,-\xi I_8),
\label{eq:4.33}
\end{equation}
where $\xi=e^{2\pi i/16}$ and $\gamma_{1,6}^8=\gamma_{1,6'}^8=-1$.

For the case of Eq.(\ref{eq:4.8}), ${\rm Tr}\gamma^{-1}_{\Omega Rk,6}\gamma^T_{\Omega Rk,6} = - {\rm Tr}\gamma^{-1}_{\Omega Rk,6'}\gamma^T_{\Omega Rk,6'} = -32$ and $\gamma_{1,p}, p=6,6'$ is simply obtained by multiplying the phase factor $e^{i \pi/4}$ to Eq.(\ref{eq:4.33}).

The $\gamma$ matrix (\ref{eq:4.33}) defines the shift vector
\begin {equation}
V_{66}=V_{6'6'}=\frac{1}{16}(\underbrace{1\cdots 1}_{8},
\underbrace{7\cdots7}_{8})
\label{eq:4.34}
\end{equation}
which selects the gauge group by the condition (\ref{eq:2.19}) as
\begin{equation}
U(8)^2 \times U(8)^2.
\label{eq:4.35}
\end{equation}
Chiral charged matter fields are obtained by the condition (\ref{eq:2.20}). The 66 states are given by
\begin{equation}
(28,1;1,1), ~~(1,\overline{28};1,1),~~(8,\bar{8};1,1)
\label{eq:4.36}
\end{equation}
and the $6'6'$ states are 
\begin{equation}
\begin{array}{rrr}
(1,1;28,1),& (1,1;1,\overline{28}),& (1,1;8,\bar{8}).
\end{array}
\label{eq:4.37}
\end{equation}
The $66'$ states are determined by Eq.(\ref{eq:2.22}) as follows,
\begin{equation}
({\bar 8},1;{\bar 8},1),~~(1,8;1,8)
\label{eq4.38}
\end{equation}

(5) $Z'_{12}$

In this model there are tadpole divergences from ${Z}_{pq}(\theta^k), p,q = 6,6'; k=2,4,6,8,10$, and ${Z}_p(\theta^k), p = 6,6'; k = 2,4,8,10$, which are proportional to $L_8/L_9$. Cancellation of these divergences requires the following condition,
\begin {eqnarray}
&&\frac{1}{4}\sum_{p=6,6'}\left\{\sum_{k=2,10}({\rm Tr}\gamma_{k,p})^2 + 3 \sum_{k=4,8}({\rm Tr}\gamma_{k,p})^2 + 4({\rm Tr}\gamma_{6,p})^2 \right\}
\nonumber\\
&& -\frac{1}{2}\sum_{n=1,\cdots,5}(-1)^n({\rm Tr}\gamma_{2n,6})({\rm Tr}\gamma_{2n,6'})
\nonumber\\
&& - 4\sum_{k=2,10}\left\{{\rm Tr}(\gamma_{\Omega Rk,6}^{-1}\gamma_{\Omega Rk,6}^T) - {\rm Tr}(\gamma_{\Omega Rk,6'}^{-1}\gamma_{\Omega Rk,6'}^T)\right\}
\nonumber\\
&& - 12\sum_{k=4,8}\left\{{\rm Tr}(\gamma_{\Omega Rk,6}^{-1}\gamma_{\Omega Rk,6}^T) - {\rm Tr}(\gamma_{\Omega Rk,6'}^{-1}\gamma_{\Omega Rk,6'}^T)\right\}
=0.  
\label{eq:4.39}
\end{eqnarray} 
In Eq.(\ref{eq:4.39}), ${\rm Tr}(\gamma_{\Omega Rk,6}^{-1}\gamma_{\Omega Rk,6}^T) = -{\rm Tr}(\gamma_{\Omega Rk,6'}^{-1}\gamma_{\Omega Rk,6'}^T) = 32$ due to Eq.(\ref{eq:4.7}) or Eq.(\ref{eq:4.8}).

The divergences from ${Z}_{p,q}(\theta^k), p,q =6,6'; k= 1,3,5,7,9,11$ are canceled by the condition given by,
\begin{eqnarray}
&& \sum_{p=6,6'}\left\{\sum_{k=1,5,7,11}({\rm Tr}\gamma_{k,p})^2 + 2\sum_{k=3,9}({\rm Tr}\gamma_{k,p})^2 \right\}
\nonumber\\
&& + 2\sum_{k=1,5,7,11}({\rm Tr}\gamma_{k,6})({\rm Tr}\gamma_{k,6'}) - 2\sum_{k=3,9}({\rm Tr}\gamma_{k,6})({\rm Tr}\gamma_{k,6'}) = 0.
\label{eq:4.40}
\end{eqnarray}
These two conditions (\ref{eq:4.39}) and (\ref{eq:4.40}) are satisfied by the following $\gamma$ matrices,
\begin{equation}
\begin{array}{ll}
{\rm Tr}\gamma_{k,6} = {\rm Tr}\gamma_{k,6'} = 0, & k \neq 4,8 \\
{\rm Tr}\gamma_{4,p} = {\rm Tr}\gamma_{8,p} = 32, & p = 6, 6'
\end{array}
\label{eq:4.41}
\end{equation}
where $\gamma_{1,p}$ is given by
\begin {equation}
\gamma_{1,6}=\gamma_{1,6'}
= \mbox{diag}(\zeta^3I_8,-I_4,I_4,\zeta^9I_8,-I_4,I_4)
\label{eq:4.42}
\end{equation} 
with $\zeta=e^{2\pi i/12}$. This $\gamma$ matrix (\ref{eq:4.42}) defines the shift vector,
\begin {equation}
V_{66}=V_{6'6'}=\frac{1}{12}(\underbrace{3\cdots 3}_{8},
\underbrace{6\cdots6}_{4},\underbrace{0\cdots0}_{4}).
\label{eq:4.43}
\end{equation}

The gauge symmetry is determined by Eq.(\ref{eq:2.19}) as
\begin{equation}
[U(8)\times SO(8)^2] \times [U(8)\times Sp(8)^2].
\label{eq:4.44}
\end{equation}
The charged chiral 66 states are given by
\begin{equation}
(28,1,1;1^3), ~~(\overline{28},1,1;1^3), ~~(1,8,8;1^3)
\label{eq:4.45}
\end{equation}
and the $6'6'$ states amount to
\begin{equation}
\begin{array}{rrr}
(1^3;28,1,1),& (1^3;\overline{28},1,1),& (1^3;1,8,8)\\
(1^3;8,1,1),& (1^3;\bar{8},1,1). & \mbox{ }
\end{array}
\label{eq:4.46}
\end{equation}
Finally, the $66'$ states are given by
\begin{equation}
\begin{array}{rr}
(8,1,1;1,1,8),& ({\bar 8},1,1;1,8,1) \\
(1,8,1;{\bar 8},1,1),& (1,1,8;8,1,1)
\end{array}
\label{eq:4.47}
\end{equation}

The remaining $Z_N$ orientifold, i.e., $Z_6, Z'_6$ and $Z_{12}$ are studied in a similar way. It turns out that tadpoles of these orientifolds cannot be canceled by simply introducing 6- and $6'$-branes.

(6) $Z_4 \times Z_4$

We consider $Z_4 \times Z_4$ with the twist vectors $v_{\theta} = \frac{1}{4}(1,-1,0)$ and $v_{\omega} = \frac{1}{4}(0,1,-1)$. As noted in Sec.2 there is a set of $6_1, 6_2, 6_3$-branes, where $6_1$-branes have the D condition for $X_5,X_6,X_8$ and $6_2$-branes have the D condition for $X_4,X_7,X_8$. The $6_3$-branes are denoted by $6'$-branes up till now.
The tadpole singularity proportional to $L_4 L_6 L_8/L_5 L_7 L_9$ is canceled by introducing 6-branes and the one proportional to $L_5 L_7 L_8/L_4 L_6 L_9$ is canceled by $6_3$-branes as usual. The new feature is the appearance of the divergences proportional to $L_4 L_7 L_9/L_5 L_6 L_8$ and $L_5 L_6 L_9/L_4 L_7 L_8$. These divergences are canceled by $6_1$ and $6_2$-branes, respectively.

The divergences from the cylinder and M\"obius strip amplitudes are canceled by taking the following $\gamma$ matrices.
\begin{equation}
\begin{array}{lcl}
\gamma_{\omega,6}&=& \gamma_{\omega,6_3} = \gamma_{\theta,6_2} = \mbox{diag}(I_8,-I_8,I_8,-I_8)
\nonumber \\
\gamma_{\theta,6}&=& \mbox{diag}(I_4,-I_4,I_4,-I_4,I_4,-I_4,I_4,-I_4)
\nonumber\\
\gamma_{\theta,6_3}&=& \gamma_{\omega,6_2} = \mbox{diag}(I_2,-I_2,\epsilon I_4,I_2,-I_2,\epsilon 
I_4,I_2,-I_2,\epsilon^3 I_4,I_2,-I_2,\epsilon^3 I_4)
\nonumber\\
\gamma_{\omega,6_1}&=& \mbox{diag}(I_4,-I_4,\epsilon I_8,I_4,-I_4,\epsilon^3 
I_8)
\nonumber\\
\gamma_{\theta,6_1}&=& \mbox{diag}(\epsilon I_2,\epsilon^3 I_2,\epsilon 
I_2,\epsilon^3 I_2,I_4,-I_4,\epsilon I_2,\epsilon^3 I_2,\epsilon 
I_2,\epsilon^3 I_2,I_4,-I_4,)
\end{array}
\label{eq:4.48}
\end{equation}
where $\epsilon = e^{2\pi i/4}$. The corresponding shift vectors are given by
\begin{equation}
\begin{array}{lcl}
V_{\omega,6}&=& V_{\omega,6_3} = V_{\theta,6_2} = \frac{1}{4}(\underbrace{0\cdots 
0}_{8},\underbrace{2\cdots2}_{8})
\nonumber \\
V_{\theta,6}&=&\frac{1}{4}(\underbrace{0\cdots 
0}_{4},\underbrace{2\cdots2}_{4},
\underbrace{0\cdots 0}_{4},\underbrace{2\cdots2}_{4})
\nonumber \\
V_{\theta,6_3}&=& V_{\omega,6_2} = \frac{1}{4}(0,0,2,2,\underbrace{1\cdots 
1}_{4},0,0,2,2,\underbrace{1\cdots 1}_{4})
\nonumber \\
V_{\omega,6_1}&=&\frac{1}{4}(\underbrace{0\cdots 
0}_{4},\underbrace{2\cdots2}_{4},
\underbrace{1\cdots 1}_{8})
\nonumber \\
V_{\theta,6_1}&=&\frac{1}{4}(1,1,3,3,1,1,3,3,\underbrace{0\cdots 
0}_{4},\underbrace{2\cdots2}_{4})
\end{array}
\label{eq:4.49}
\end{equation}
The gauge symmetry of this orientifold is determined by $\rho\cdot V = 0 \mbox{ mod } {\bf Z}$, where $\rho$ is the root vectors of $SO(32)$ for 6-branes and $Sp(32)$ for $6_{1,2,3}$-branes. Using the shift vectors given by Eq.(\ref{eq:4.49}), we obtain the gauge group of the $Z_4 \times Z_4$ orientifold as follows,
\begin{equation}
SO(8)^4 \times Sp(4)^8 \times U(4)^6 \times U(2)^4,
\label{eq:4.50}
\end{equation}
where $SO(8)^4$ comes from 6-branes, $Sp(4)^4 \times U(4)^2$ is from $6_3$-branes, $U(4)^2 \times U(2)^4$ is from $6_1$-branes and $Sp(4)^4 \times U(4)^2$ comes from $6_2$-branes.
%
\section{Discussions}

We have constructed the $D = 4, N =1$, type IIA orientifolds for $Z_N$ that act crystallographically on a $T^6$ lattice. We found that among the possible $Z_N$, $Z_4, Z_8, Z'_8, Z'_{12}$ are consistent orientifolds while $Z_3, Z_7, Z_6, Z'_6, Z'_{12}$ are inconsistent due to the impossibility of tadpole cancellation. This result is reverse to what was obtained in $D = 4, N = 1$ type IIB orientifolds, where $Z_3, Z_7, Z_6, Z'_6, Z'_{12}$ are consistent type IIB orientifolds but $Z_4, Z_8, Z'_8, Z'_{12}$ are not. In type IIB orientifolds, the Klein bottle amplitude for $Z_4, Z_8, Z'_8, Z'_{12}$ has divergences proportional to $V_4/L_8L_9$ that cannot be canceled against any of the M\"obius strip or cylinder contributions \cite{ald}. On the other hand, in type IIA orientifolds all divergences in the Klein bottle amplitude are canceled by introducing 32 6- and $6'$-branes and there is no further divergence in this amplitude. 

The different behavior of the type IIA and type IIB Klein bottle amplitudes for $t \rightarrow 0$ is due to the difference of the boundary conditions in the closed string sector. For type IIB closed strings, the orbifold action $\theta$ acts on the $\Omega$ invariant states as
\begin{equation}
\theta^k Y\tilde{Y}|0\rangle = e^{4\pi ikv_i}Y\tilde{Y}|0\rangle .
\label{eq:5.1}
\end{equation}
On the other hand, $\theta$ acts on the $\Omega R$ invariant states of type IIA closed strings as
\begin{equation}
\theta^k Y \bar{\tilde{Y}}|0\rangle = Y\bar{\tilde{Y}}|0\rangle .
\label{eq:5.2}
\end{equation}
Thus the type IIA Klein bottle amplitude behaves as Eq.(\ref{eq:3.4}) or Eq.(\ref{eq:3.5}), hence it does not give tadpole divergence if $2kv_i \neq$ integer. When $2kv_i =$ integer, the divergent contribution from the tadpole diagram is canceled by introducing 6-/$6'$-branes.

Therefore, in the type IIA orientifolds extra divergences which arise from the M\"obius strip and cylinder amplitudes must be canceled by themselves. As explained in Sec.4, this cancellation is impossible for $Z_3$ and $Z_7$. For other orientifolds, i.e., $Z_6, Z'_6, Z_{12}$, similar analysis shows that the tadpole divergences are not canceled. For example, in the $Z_6$ orientifold there are tadpole divergences from the cylinder amplitudes, $Z_{66}(\theta^3), Z_{6'6'}(\theta^3)$ and $Z_{66'}(\theta^3)$ which are proportional to $L_8/L_9$. The cancellation condition is given by
\begin{equation}
2({\rm Tr}\gamma_{3,6})^2 + 2({\rm Tr}\gamma_{3,6'})^2 + {\rm Tr}\gamma_{3,6}{\rm Tr}\gamma_{3,6'} = 0.
\label{eq:5.3}
\end{equation}
There are also tadpole divergences from $Z_{pq}(\theta^k), Z_p(\theta^k), p, q = 6, 6'$ and $k = 1,2,4,5$ which have no six dimensional compact volume dependence. The cancellation condition for these divergences amount to
\begin{eqnarray}
&&\sum_{p=6,6'}\left\{\sum_{k=1,5}\left[\frac{1}{4}({\rm Tr}\gamma_{k,p})^2 - 2{\rm Tr}(\gamma^{-1}_{\Omega Rk,p}\gamma^{T}_{\Omega Rk,p})\right]\right.
\nonumber\\
&&~~~~~\left.+ \sum_{k=2,4}\left[\frac{3}{4}({\rm Tr}\gamma_{k,p})^2+ 6{\rm Tr}(\gamma^{-1}_{\Omega Rk,p}\gamma^{T}_{\Omega Rk,p})\right]\right\} 
\nonumber\\
&& ~~~~~~~~~~+ \frac{1}{2}\sum_{k=1,2,4,5}({\rm Tr}\gamma_{k,6})({\rm Tr}\gamma_{k,6'}) = 0.
\label{eq:5.4}
\end{eqnarray}
It turns out that there are no $\gamma_{1,6}$ and $\gamma_{1,6'}$ which satisfy these tadpole cancellation conditions.

Let us consider the impossibility of tadpole cancellation for $Z_3, Z_7, Z_6, Z'_6, Z'_{12}$ from a somewhat different viewpoint.
In the type IIA $Z_N$ orientifolds with 6- and $6'$-brane configuration, each complex plane has two target space coordinates and the corresponding string wave functions have the Neumann and the Dirichlet boundary conditions.
In fact, 6-branes have the Neumann condition for $X_4, X_6, X_8$ and the Dirichlet condition for $X_5, X_7, X_9$. As for $6'$-branes, they have the Neumann condition for $X_5, X_7, X_8$ and the Dirichlet condition for $X_4, X_6, X_9$. So, each complex plane $Y_i$ has the string wave functions with the Neumann and the Dirichlet conditions. Under the $Z_N$ rotation, these two string wave functions with the different boundary conditions are mixed. However, by rotating the coordinate axes we can redefine the $Z_N$ action as
\begin{eqnarray}
\theta' &=& \exp(2i\pi(v'_1 J_{46} + v'_2 J_{57} + v_3 J_{89}))
\nonumber\\
&=& U \exp(2i\pi(v_1 J_{45} + v_2 J_{67} + v_3 J_{89})) U^{-1},
\label{eq:5.5}
\end{eqnarray}
where $U$ is an appropriate $SO(6)$ transformation. Then $\theta'$ acts diagonally on $Y'_1 = X_4 + iX_6$ and $Y'_2 = X_5 + iX_7$. This time, both $X_4$ and $X_6$ in the $Y'_1$ plane have the same type of the boundary condition. Similar situation holds for $X_5$ and $X_7$ in the $Y'_2$ plane. In this way we can rearrange the two complex planes such that there is no mixing between the Neumann and the Dirichlet conditions under the $\theta'$ rotation. 

For the third plane $Y_3$, however, we cannot get rid of the mixing of the two boundary conditions as we show in the following. Brane configurations of the $Z_N$ orientifold should be invariant under the $Z_N$ rotation. In particular, under $Y'_3 = \exp(2\pi ikv_3)Y_3$ 
\begin{equation}
(\partial_\sigma' X'_8, \partial_\tau' X'_9)
= (\partial_\sigma X_8, \partial_\tau X_9),
\label{eq:5.6}
\end{equation}
so that the Neumann condition $\partial_\sigma X_8 = \partial_\sigma' X'_8 = 0$ and the Dirichlet condition $\partial_\tau X_9 = \partial_\tau' X'_9 = 0$ at $\sigma, \sigma' = 0, \pi$ are kept invariant under the $Z_N$ rotation of the 8-9 plane.
This implies that the world-sheet coordinates must be transformed under $Z_N$ in such a way as
\begin{equation}
\left(
\begin{array}{c}
\partial_\sigma' \\
\partial_\tau'
\end{array}
\right)
=
\left(
\begin{array}{rr}
\cos 2\pi kv_3 & -\sin 2\pi kv_3 \\
\sin 2\pi kv_3 & \cos 2\pi kv_3
\end{array}
\right)
\left(
\begin{array}{c}
\partial_\sigma \\
\partial_\tau
\end{array}
\right)
\label{eq:5.7}
\end{equation}
Then, consistency with the $Z_N$ invariance requires that the open strings with the rotated boundary conditions should be included. The rotated boundary conditions are written by
\begin{eqnarray}
\cos 2\pi kv_3 \partial_\sigma X_8 - \sin 2\pi kv_3 \partial_\tau X_8 &=& 0,
\label{eq:5.8}\\
\cos 2\pi kv_3 \partial_\tau X_9 + \sin 2\pi kv_3 \partial_\sigma X_9 &=& 0,
\label{eq:5.9}
\end{eqnarray}
at $\sigma = 0, \pi$. 
When 2$kv_3 = $ integer, Eq.(\ref{eq:5.8}) reduces to the Neumann condition while Eq.(\ref{eq:5.9}) gives the Dirichlet condition. When 2$kv_3 = $ half integer, Eq.(\ref{eq:5.8}) reduces to the Dirichlet condition and Eq.(\ref{eq:5.9}) turns out to the Neumann condition. For 4$kv_3 \neq$ integer, we obtain the mixed boundary condition which is neither Neumann nor Dirichlet. 

Appearance of the sectors with the mixed boundary conditions or twisted open strings in the type IIB superstring has been discussed in Refs.\cite{kst,blum}.  As in the type IIB orientifolds, the open strings with mixed boundary conditions do not end on the D-branes. The endpoint of such open strings is not stuck on a rigid manifold but rather it harmonically oscillates around a fixed point. There is no consistent world-sheet, {\it i.e.}, perturbative description of these phenomena within the orientifold approach \cite{kst}.

The type IIA orientifolds of $Z_3, Z_7, Z_6, Z_{12}$ have the twist vector with 4$kv_3 \neq$ integer for $k \neq 0, \frac{N}{2}$ and they have open strings with the mixed boundary condition. So we expect that although tadpole cancellation conditions are not satisfied by these orientifolds perturbatively, there may be additional non-perturbative contributions from these open strings. For $Z'_6$, a little bit careful analysis is needed. Since $v_2 = -\frac{1}{2}$ in this model, we change the coordinate axes such that
\begin{equation}
\theta' = \exp(2i\pi(v'_1 J_{48} + v_2 J_{67} + v'_3 J_{59}))
\label{eq:5.10}
\end{equation}
For this choice of the axes, the 6-brane configuration has the Neumann condition in the 4-8 plane and the Dirichlet condition in the 5-9 plane. On the other hand, the $6'$-brane configuration has the mixed boundary condition in the 4-8 and 5-9 planes. It is impossible to choose coordinate axes to eliminate mixed boundary conditions for both 6- and $6'$-branes simultaneously. Thus the $Z'_6$ orientifolds have non-perturbative sectors and do not satisfy the perturbative tadpole cancellation condition.

For the type IIA orientifolds of $Z_4, Z_8, Z'_8, Z'_{12}$, the twist vectors obey $2v_3$ = integer or half-integer so that the boundary conditions (\ref{eq:5.8}) and (\ref{eq:5.9}) are the Neumann or the Dirichlet. Thus we can always choose the coordinate axes so as to eliminate the mixed boundary conditions and these orientifolds obey the perturbative tadpole cancellation conditions.

To summarize we have derived the tadpole cancellation conditions for the type IIA $Z_N$ orientifolds and found that the $Z_4, Z_8, Z'_8, Z'_{12}$ orientifolds satisfy the tadpole cancellation conditions while the $Z_3, Z_7, Z_6, Z'_6, Z_{12}$ orientifolds do not. In Table 2 we summarize the gauge group and charged chiral multiplets of the $Z_4, Z_8, Z'_8, Z'_{12}$ orientifolds. Extension of our argument to $Z_N \times Z_M$ orientifolds is straightforward and we gave an example of the $Z_4 \times Z_4$ orientifold.
We have argued that for the $Z_3, Z_7, Z_6, Z'_6, Z_{12}$ orientifolds, there exist open strings with the mixed boundary conditions and their end-points are not D-branes but some non-perturbative objects. Appearance of the non-perturbative sector could be the reason why these orientifolds do not obey the tadpole cancellation conditions obtained perturbatively. This result is just opposite of what was obtained in the type IIB orientifolds, where the $Z_3, Z_7, Z_6, Z'_6, Z_{12}$ orientifolds obey the perturbative tadpole cancellation conditions while the $Z_4, Z_8, Z'_8, Z'_{12}$ orientifolds do not. Here, the latter orientifolds have the non-perturbative sector from the open strings with the mixed boundary conditions \cite{kst}. This implies that the perturbatively consistent vacua are changed into the non-perturbative vacua under the T-duality between the type IIA and the type IIB orientifolds in four dimensions.
%
%
\begin{table}
\begin{center}
\begin{tabular}{|c|c|c|c|}\hline
Twist Group & Gauge Group & (66)/($6'6'$) matter & (6$6'$) matter \\
\hline
$Z_4$ & 
\begin{tabular}{r}
$[U(8) \times SO(8)^2]$ \\
$\times [U(8) \times Sp(8)^2]$
\end{tabular}  & 
\begin{tabular}{rr}
2(8,1,8;$1^3$) & 2($\bar{8}$,8,1;$1^3$) \\
(28,1,1;$1^3$) & ($\overline{28}$,1,1;$1^3$) \\
(1,8,8;$1^3$) & \\
\hline
2($1^3$;8,1,8) & 2($1^3$;$\bar{8}$,8,1) \\
($1^3$;28,1,1) & ($1^3$;$\overline{28}$,1,1)\\
($1^3$;8,1,1) & ($1^3$;$\bar{8}$,1,1) \\
($1^3$;1,8,8) & 
\end{tabular} &
\begin{tabular}{r}
(8,1,1;1,1,8) \\ (1,1,8;8,1,1) \\
(1,8,1;${\bar 8}$,1,1) \\ (${\bar 8}$,1,1;1,8,1)
\end{tabular} \\
\hline
$Z_8$ & 
\begin{tabular}{r}
$[U(8) \times SO(8)^2]$ \\
$\times [U(4)^3 \times Sp(4)^2]$
\end{tabular} &
\begin{tabular}{rr}
(28,1,1;$1^5$) & ($\overline{28}$,1,1;$1^5$) \\
(1,8,8;$1^5$) & $\mbox{ }$ \\
\hline
($1^3$;4,1,1,1,4) & ($1^3$;4,4,1,1,1) \\
($1^3$;$\bar{4}$,4,1,1,1) & ($1^3$;1,$\bar{4}$,4,1,1) \\
($1^3$;1,$\bar{4},\bar{4}$,1,1) & ($1^3$;1,1,$\bar{4}$,4,1) \\
($1^3$;1,1,4,1,4) & ($1^3$;$\bar{4}$,1,1,4,1) \\
($1^3$;1,1,1,4,4) & ($1^3$;4,1,4,1,1) \\
($1^3$;$\bar{4}$,1,$\bar{4}$,1,1) & ($1^3$;1,4,1,1,1) \\
($1^3$;1,$\bar{4}$,1,1,1) & ($1^3$;1,6,1,1,1) \\
($1^3$;1,$\bar{6}$,1,1,1) & $\mbox{ }$
\end{tabular} &
\begin{tabular}{r}
(8,1,1;1,1,1,1,4) \\ (${\bar 8}$,1,1;1,1,1,4,1) \\
(1,8,1;1,4,1,1,1) \\ (1,1,8;1,${\bar 4}$,1,1,1)
\end{tabular} \\
\hline
$Z'_8$ & $U(8)^2 \times U(8)^2$ &
\begin{tabular}{rr}
(28,1;1,1) & (1,$\overline{28}$;1,1)\\
(8,$\bar{8}$;1,1) & $\mbox{ }$\\
\hline
(1,1;28,1) & (1,1;1,$\overline{28}$) \\
(1,1;8,$\bar{8}$) & \mbox{ }
\end{tabular} &
\begin{tabular}{r}
({$\bar 8$},1;{$\bar 8$},1) \\ (1,8;1,8)
\end{tabular} \\
\hline
$Z'_{12}$ & 
\begin{tabular}{r}
$[U(8) \times SO(8)^2]$ \\
$\times [U(8) \times Sp(8)^2]$ 
\end{tabular} &
\begin{tabular}{rr}
(28,1,1;$1^3$) & ($\overline{28}$,1,1;$1^3$) \\
(1,8,8;$1^3$) & $\mbox{ }$ \\
\hline
($1^3$;28,1,1) & ($1^3$;$\overline{28}$,1,1) \\
($1^3$;1,8,8) & ($1^3$;8,1,1) \\
($1^3$;$\bar{8}$,1,1) & $\mbox{ }$
\end{tabular} &
\begin{tabular}{rr}
(8,1,1;1,1,8) \\ (${\bar 8}$,1,1;1,8,1) \\
(1,8,1;{$\bar 8$},1,1) \\ (1,1,8;8,1,1)
\end{tabular} \\
\hline
\end{tabular}
\end{center}
\caption{Gauge group and charged chiral multiplets in the type IIA $Z_N$ orientifolds.}
\end{table}
\newpage
%
\appendix
\section{Appendix}

{\bf One-loop amplitudes of the type IIA orientifolds}

In this appendix we give the detailed expression for the one-loop amplitudes of the Klein bottle$({\cal K})$, the M\"{o}bius strip$({\cal M})$, and the cylinder$({\cal C})$.

(1) Klein bottle amplitude

The Klein bottle amplitude is given by Eqs.(\ref{eq:3.1}) and (\ref{eq:3.2}), where $L_0(\theta^n)$ and $\tilde {L}_0(\theta^n)$ are defined by
\begin {equation}
\begin{array}{lcl}
L_0(\theta^n) &=&\frac{1}{2}(\frac{P}{2}-L)^2+N(\theta^n)+\frac{1}{8}P_\mu^2+a \\
\tilde{L}_0(\theta^n) &=&\frac{1}{2}(\frac{P}{2}+L)^2+\tilde{N}(\theta^n)
+\frac{1}{8}P_\mu^2+\tilde{a}
\end{array}
\label{eq:6.1}
\end{equation}
where $P_{\mu}$ is an uncompactified four-dimensional momentum and $P, L$ are compactified internal momenta and windings. $N(\theta^n)$ and $\tilde{N}(\theta^n)$ are the oscillator part of the left-moving and right-moving Hamiltonian, respectively. $a (\tilde{a})$ is a left (right)-moving normal-ordering constant.

The trace in $Z_{\cal K}$ is computed in a standard way. At first we represent $L_0(\theta^n)={\cal L}_1+{\cal L}_2+{\cal L}_3$ where 
\begin{equation}
{\cal L}_1=\frac{1}{8}P_\mu^2, \mbox{ }{\cal L}_2=N(\theta^n)+a, \mbox{ }{\cal L}_3=\frac{1}{2}(\frac{P}{2}-L)^2
\label{eq6:2}
\end{equation}
and similarly for $\tilde{L}_0(\theta^n)$.
Then we find that the contribution of ${\cal L}_1$ and $\tilde{{\cal L}_1}$ is given by 
\begin {equation}
Z_{\cal K}^{(1)}(\theta^n,\theta ^k)=\frac{1}{(2\pi ^2t)^2},
\label{eq:6.3}
\end{equation}
As for ${\cal L}_2$ and $\tilde{{\cal L}_2}$ there are contributions from both the untwisted sector($n=0$) and the twisted sector($n \ne 0$).
For the untwisted sector we obtain
\begin {eqnarray}
Z_{\cal K}^{(2)}(1,\theta ^k)
&=&\frac{1}{2\tilde{\eta}^{12}}
\Bigl[\tilde{\vartheta}[^0_0]^4-\tilde{\vartheta}[^0_{\frac{1}{2}}]^4
-\tilde{\vartheta}[^{\frac{1}{2}}_0]^4 \Bigr] \nonumber \\
&=&(1-1)\frac{1}{2\tilde{\eta}^{12}}\tilde{\vartheta}[^0_{\frac{1}{2}}]^4
\label{eq:6.4}
\end{eqnarray} 
where the $\theta$ function and the Dedekind $\eta$ function are defined respectively by
\begin {eqnarray}
\vartheta[^{\delta}_{\varphi}](t)&=&\sum_{n}q^{\frac{1}{2}(n+\delta)^2}
e^{2i\pi (n+\delta )\varphi}
\label{eq:6.5}\\
\eta&=&q^{\frac{1}{24}}\prod_{n=1}^{\infty}(1-q^n)
\label{eq:6.6}
\end{eqnarray}
Here $q=e^{-2\pi t}$ and $\tilde{\vartheta},\tilde{\eta}$ indicate functions of $\tilde{q}=q^2=e^{-4\pi t}$ instead of $q$.  In deriving Eq.(\ref{eq:6.4}) we have used the Riemann identities for the $\theta$ functions \cite{mum}.
To extract the divergences, we take the limit t $\to$ 0 and obtain
\begin {equation}
Z^{(2)}_{\cal K}(1,\theta^k)\to (1-1)\frac{1}{2}(4t)^4
\label{eq:6.7}
\end{equation}

For the twisted sector, only $n=\frac{N}{2}$ for $N$= even, survives and
\begin {eqnarray}
Z^{(2)}_{\cal K}(\theta^{\frac{N}{2}},\theta ^k)
&=&\frac{1}{2\tilde{\eta}^4}\tilde{\vartheta}[^0 _{\frac{1}{2}}]^{-2}
\Bigl[ \tilde{\vartheta}[^0_0]^2\tilde{\vartheta}[^{\frac{1}{2}}_0]^2-0
-\tilde{\vartheta}[^0_0]^2\tilde{\vartheta}[^{\frac{1}{2}}_0]^2 \Bigr] 
\nonumber \\
&=&(1-1)0.
\label{eq:6.8}
\end{eqnarray}
Thus the twisted sectors do not contribute to the amplitude.

For ${\cal L}_3$ and $\tilde{{\cal L}_3}$, we put $P_{\mu}=n_{\mu} /R_0, L_{\mu}=m_{\mu}R_0, \rho=2R_0^2$, where $R_0$ is a common radius of the compact space. 
When $kv_i$=integer, the state $|n_{\mu},m_{\mu}\rangle$ transforms as
\begin {equation}
\Omega R \theta^k |n_{\mu},m_{\mu}\rangle=\cases{|-n_{\mu},m_{\mu}\rangle \quad (\mu=5,7,9)\cr |n_{\mu},-m_{\mu}\rangle \quad ~(\mu=4,6,8)\cr}
\label{eq:6.9}
\end{equation}
Then we obtain
\begin {eqnarray}
Z^{(3)}_{\cal K}(1,\theta ^k) &=& \prod_i \frac{1}{t}\frac{L_{e_i}}{L_{o_i}}
\sum_s e^{-\pi \rho s^2/t} \sum_s e^{-\pi s^2/t\rho} \nonumber\\
&\rightarrow& \prod_i \frac{1}{t}\frac{L_{e_i}}{L_{o_i}}
\label{eq:6.10}
\end{eqnarray}
where $L_{e_i} (L_{o_i})$ denotes the length of $\mu$=even (odd) direction of the $i$-th torus. For the sake of simplicity, we take $L_{e_i} = L_{o_i}= R_0$.
When $kv_i$=half integer, the state $|n_{\mu},m_{\mu}\rangle$ transforms as
\begin {equation}
\Omega R \theta^k |n_{\mu},m_{\mu}\rangle=\cases{|n_{\mu},-m_{\mu}\rangle \quad ~(\mu=5,7,9)\cr |-n_{\mu},m_{\mu}\rangle \quad (\mu=4,6,8)\cr}
\label{eq:6.11}
\end{equation} 
and we obtain
\begin {eqnarray}
Z^{(3)}_{\cal K}(1,\theta ^k) &=& \prod_i \frac{1}{t}\frac{L_{o_i}}{L_{e_i}}
\sum_s e^{-\pi \rho s^2/t} \sum_s e^{-\pi s^2/t\rho} \nonumber\\
&\rightarrow& \prod_i \frac{1}{t}\frac{L_{o_i}}{L_{e_i}}
\label{eq:6.12}
\end{eqnarray}

(2) Cylinder amplitude

Next we compute cylinder amplitude given by Eqs.(\ref{eq:3.6}) and (\ref{eq:3.7}), where the Virasoro operator $L_0$ is defined by
\begin {equation} 
L_0 = \frac{1}{2}(P-L)^2+N(\theta^k)+\frac{1}{2}P_\mu^2+a
\label{eq:6.13}
\end{equation}
Again we represent $L_0 ={\cal L}_1+{\cal L}_2+{\cal L}_3$ 
where
 ${\cal L}_1=\frac{1}{2}P_\mu^2,
{\cal L}_2=N(\theta^k)+a,
{\cal L}_3=\frac{1}{2}(\frac{P}{2}-L)^2 $.

The contribution from ${\cal L}_1$ is given by
\begin {equation}
Z^{(1)}_{pq}(\theta ^k)=\frac{1}{(4\pi ^2t)^2},
\label{eq:6.14}
\end{equation}

Next we consider the contribution of ${\cal L}_2$ to $Z_{66},Z_{6'6'}$. We find that
\begin{eqnarray}
Z^{(2)}_{66} (\theta^k)=Z^{(2)}_{6'6'} (\theta^k)
=\frac{1}{2\eta ^3}\prod_{i=1}^3(-2\sin \pi kv_i)
\vartheta [^{\frac{1}{2}}_{\frac{1}{2}+kv_i}]^{-1} \times \nonumber \\
\left\{ \vartheta [^0_0] \prod_{i=1}^3 \vartheta [^0_{kv_i}]
-\vartheta[^0_{\frac{1}{2}}]\prod_{i=1}^3\vartheta[^0_{\frac{1}{2}+kv_i}]
-\vartheta[^{\frac{1}{2}}_0]\prod_{i=1}^3\vartheta[^{\frac{1}{2}}_{kv_i}] \right\}
\nonumber \\
=(1-1)\frac{1}{2\eta ^3} \vartheta [^0_{\frac{1}{2}}]
\prod_{i=1}^3(-2\sin\pi kv_i)
\vartheta[^{\frac{1}{2}}_{\frac{1}{2}+kv_i}]^{-1}         \vartheta[^0_{\frac{1}{2}+kv_i}] .
\label{eq6.15}
\end{eqnarray} 
To extract the divergence, we take the limit $t \to 0$,
\begin{eqnarray}
Z^{(2)}_{66}(\theta^k)=Z^{(2)}_{6'6'}(\theta^k) &\to& (1-1)\frac{1}{2}(2t)\prod_i2t 
\prod_j(-2\sin\pi kv_j) 
\nonumber\\
&& \times \frac{(1+q_{(1/2t)}^{-kv_j})}{(1-q_{(1/2t)}^{-kv_j})} \prod_{n=1}^{\infty}
\frac{(1+q_{(1/2t)}^{n-kv_j})(1+q_{(1/2t)}^{n+kv_j})}
{(1-q_{(1/2t)}^{n-kv_j})(1-q_{(1/2t)}^{n+kv_j})}
\label{re:6.16}
\end{eqnarray}
where $kv_i$=integer and $kv_j \ne$ integer.
Here $q_{(1/2t)}$ stands for $q$ as a function of $1/2t$ insted of $t$ so that $q_{(1/2t)} \to 0$ for $t \to 0$.

The contribution of ${\cal L}_2$ to $Z_{66'}$ amounts to
\begin{eqnarray}
Z^{(2)}_{66'} (\theta^k)
&=& \frac{1}{2\eta ^3}(-2\sin \pi kv_3)
\vartheta [^{\frac{1}{2}}_{\frac{1}{2}+kv_3}]^{-1}\prod_{i=1,2}\vartheta [^0_{\frac{1}{2}+kv_i}]^{-1}
 \times \nonumber \\
&&\left\{ \vartheta [^0_0]\vartheta [^0_{kv_3}]\prod_{i=1,2}\vartheta [^{\frac{1}{2}}_{kv_i}]
-\vartheta[^0_{\frac{1}{2}}]\vartheta[^0_{\frac{1}{2}+kv_3}]
\prod_{i=1,2}\vartheta[^{\frac{1}{2}}_{\frac{1}{2}+kv_i}]
-\vartheta[^{\frac{1}{2}}_0]\vartheta[^{\frac{1}{2}}_{kv_3}]\prod_{i=1,2}\vartheta[^0_{kv_i}]
\right\} \nonumber \\
&=& (1-1)\frac{1}{2\eta ^3} \vartheta [^0_{\frac{1}{2}}]
(-2\sin\pi kv_3)\vartheta[^0_{\frac{1}{2}+kv_3}]
\vartheta[^{\frac{1}{2}}_{\frac{1}{2}+kv_3}]^{-1}
\nonumber\\
&&\times \prod_{i=1,2}\vartheta[^{\frac{1}{2}}_{\frac{1}{2}+kv_i}]
\vartheta[^0_{\frac{1}{2}+kv_i}]^{-1}
\label{eq:6.17}
\end{eqnarray} 
The limit $t \to 0$ turns out to be
\begin {eqnarray}
Z^{(2)}_{66'}(\theta^k) &\to& (1-1)\frac{1}{2}(2t)(-2\sin\pi kv_3)\frac{(1+q_{(1/2t)}^{-kv_3})}{(1-q_{(1/2t)}^{-kv_3})}\prod_{n=1}^{\infty}
\frac{(1+q_{(1/2t)}^{n-kv_3})(1+q_{(1/2t)}^{n+kv_3})}
{(1-q_{(1/2t)}^{n-kv_3})(1-q_{(1/2t)}^{n+kv_3})}
\nonumber\\
&& \times \prod_{i=1,2}\frac{(1-q_{(1/2t)}^{-kv_i})}{(1+q_{(1/2t)}^{-kv_i})}\prod_{n=1}^{\infty}
\frac{(1-q_{(1/2t)}^{n-kv_i})(1-q_{(1/2t)}^{n+kv_i})}
{(1+q_{(1/2t)}^{n-kv_i})(1+q_{(1/2t)}^{n+kv_i})}
\label{eq:6.18}
\end{eqnarray}
for $kv_3 \neq$ integer. When $kv_3 =$ integer, the part which depends on $kv_3$ should be replaced with $2t$.

Now let us consider the contribution of ${\cal L}_3$ to $Z_{pq}$. For open strings we put $P_{\mu} = n_{\mu}/R_0, L_{\mu} = 2m_{\mu}R_0$.
The $66$-states have NN boundary conditions for $\mu=4,6,8$  and DD boundary conditions for $\mu=5,7,9$. So, there are no winding $(m_{\mu}=0)$ for $\mu=4,6,8$ and no momentum $(n_{\mu}=0)$ for $\mu=5,7,9$.
When $kv_i=$ integer, $|n,m\rangle$ transform as
$\theta^k |n,m\rangle=|n,m\rangle$.
Then we obtain
\begin {eqnarray}
Z^{(3)}_{66}(\theta ^k) &=& \prod_i \frac{1}{2t}\frac{L_{e_i}}{L_{o_i}}
\sum_s e^{-\pi \rho s^2/2t} \sum_s e^{-\pi s^2/2t\rho}
\nonumber\\
&\to& \prod_i \frac{1}{2t}\frac{L_{e_i}}{L_{o_i}}
\label{eq:6.19}
\end{eqnarray}
When $kv_i$= half integer, $|n,m\rangle$ transform as
$\theta^k |n,m\rangle=|-n,-m\rangle$
and there is no contribution to $Z_{66}$.

The $6'6'$-states have NN boundary conditions for $\mu=5,7,8$ and DD boundary conditions for $\mu=4,6,9$.
In this case there are no winding $(m_{\mu}=0)$ for $\mu=5,7,8$ 
and no momentum $(n_{\mu}=0)$ for $\mu=4,6,9$.
For $kv_i$=integer, we have
\begin {eqnarray}
Z^{(3)}_{6'6'}(\theta ^k) &=& \prod_i \frac{1}{2t}\frac{L_{m_i}}{L_{\ell_i}}
\sum_s e^{-\pi \rho s^2/2t} \sum_s e^{-\pi s^2/2t\rho}
\nonumber\\
&\to& \prod_i \frac{1}{2t} \frac{L_{m_i}}{L_{\ell_i}}
\label{eq:6.20}
\end{eqnarray} 
where $L_{m_i}=(L_5,L_7,L_8)$ and $L_{\ell_i}=(L_4,L_6,L_9)$.
For $kv_i$= half integer, $|n,m\rangle$ transform as
$\theta^k |n,m\rangle=|-n,-m\rangle$ and they do not contribute to $Z_{6'6'}$.

The $66'$-states have DN boundary conditions for $\mu=4,5,6,7$ and NN boundary condition for $\mu=8$ and DD boundary condition for $\mu=9$.
Then there are no winding $(m_{\mu}=0)$ and no momentum $(n_{\mu}=0)$ for $\mu=4,5,6,7$ 
, no winding $(m_{\mu}=0)$ for $\mu=8$ and no momentum $(n_{\mu}=0)$ for $\mu=9$.
Then, for $kv_i=$integer, we obtain
\begin {eqnarray}
Z_{66'}(\theta ^k) &=& \frac{1}{2t}\frac{L_8}{L_9}
\sum_s e^{-\pi \rho s^2/2t} \sum_s e^{-\pi s^2/2t\rho}
\nonumber\\
&\to& \frac{1}{2t}\frac{L_8}{L_9}
\label{eq:6.21}
\end{eqnarray}
For $kv_i=$ half integer, $|n,m\rangle$ transform as
$\theta^k |n,m\rangle=|-n,-m\rangle$ and they do not contribute to $Z_{66'}$.

(3) M\"{o}bius strip amplitude

Next we compute M\"{o}bius strip amplitude defined by Eqs.(\ref{eq:3.12}) and (\ref{eq:3.13}) where $L_0$ is given by Eq.(\ref{eq:6.13}). We decompose $L_0$ as in the cylinder amplitude. Then
${\cal L}_1$  contributes to the amplitude as
\begin {equation}
Z^{(1)}_p(\theta ^k)=\frac{1}{(4\pi ^2t)^2}
\label{eq:6.22}
\end{equation}

The contribution from ${\cal L}_2$ amounts to
\begin {eqnarray}
Z^{(2)}_6(\theta ^k)&=&\frac{1}{2\eta '^6}\prod_{i=1}^3(-2\sin \pi kv_i)\eta '
\vartheta '[^{\frac{1}{2}}_{\frac{1}{2}+kv_i}]^{-1}\nonumber \\
&&\times \left\{\vartheta '[^0_0] \prod_{i=1}^3 \vartheta '[^0_{kv_i}]
-\vartheta '[^0_{\frac{1}{2}}]\prod_{i=1}^3\vartheta'
[^0_{\frac{1}{2}+kv_i}]
-\vartheta '[^{\frac{1}{2}}_0]\prod_{i=1}^3\vartheta'
[^{\frac{1}{2}}_{kv_i}] \right\} \nonumber \\
&=& -(1-1)\frac{1}{2\eta '^3} \vartheta ' [_0^{\frac{1}{2}}]
\prod_{i=1}^3(-2\sin\pi kv_i)
\vartheta '[^{\frac{1}{2}}_{\frac{1}{2}+kv_i}]^{-1}
\vartheta '[^{\frac{1}{2}}_{kv_i}] 
\label{eq:6.23}
\end{eqnarray}
here $\eta'$ and $\vartheta'$ stands for the functions of $-q$ insted of $q$.
The limit $t \to 0$ of Eq.(\ref{eq:6.23}) is given by
\begin {eqnarray}
Z^{(2)}_6(\theta ^k) &\rightarrow& -(1-1)\frac{1}{2}(4t)\prod_{i} (4t)(-1)^{kv_i} 
\prod_j
(-2\sin \pi kv_j)\frac{(1+q_{(1/2t)}^{-kv_j})}{(1-q_{(1/2t)}^{-kv_j})}
\nonumber\\
&\times& \prod_{n=1}^{\infty}\frac{(1-q_{(1/2t)}^{n-kv_j-\frac{1}{2}})(1-q_{(1/2t)}^{n+kv_j-\frac{1}{2}})(1+q_{(1/2t)}^{n-kv_j})(1+q_{(1/2t)}^{n+kv_j})}{(1+q_{(1/2t)}^{n-kv_j-\frac{1}{2}})(1+q_{(1/2t)}^{n+kv_j-\frac{1}{2}})(1-q_{(1/2t)}^{n-kv_j})(1-q_{(1/2t)}^{n+kv_j})}
\label{eq:6.24}
\end{eqnarray}
for $kv_i =$ integer and $kv_j \neq$ integer.

For $6'$-branes we find
\begin {eqnarray}
Z_{6'}^{(2)}(\theta ^k)&=&\frac{1}{2\eta '^3}(-2\sin \pi kv_3)
\vartheta '[^{\frac{1}{2}}_{\frac{1}{2}+kv_3}]^{-1}
\prod_{i=1,2}2\cos \pi kv_i \> \vartheta '[^{\frac{1}{2}}_{kv_i}]^{-1}
\nonumber \\
&\times&\left\{
\vartheta '[^0_0] \vartheta '[^0_{kv_3}]
\prod_{i=1,2}\vartheta '[^0_{\frac{1}{2}+kv_i}]
-\vartheta '[^0_{\frac{1}{2}}]\vartheta '[^0_{\frac{1}{2}+kv_3}]
\prod_{i=1,2}\vartheta'[^0_{kv_i}]\right.\nonumber \\
&& ~~~~~~~~~~+ \left.\vartheta'[^{\frac{1}{2}}_0]
\vartheta'[^{\frac{1}{2}}_{kv_3}]
\prod_{i=1,2}\vartheta'[^{\frac{1}{2}}_{\frac{1}{2}+kv_i}]
\right\} \nonumber \\
&=&-(1-1)\frac{1}{2\eta '^3}\vartheta '[^{\frac{1}{2}}_0](-2\sin \pi kv_3)
\vartheta '[^{\frac{1}{2}}_{\frac{1}{2}+kv_3}]^{-1}
\vartheta'[^{\frac{1}{2}}_{kv_3}]\nonumber \\
&\times& \prod_{i=1,2}(2\cos \pi kv_i)\vartheta '[^{\frac{1}{2}}_{kv_i}]^{-1}
\vartheta'[^{\frac{1}{2}}_{\frac{1}{2}+kv_i}].
\label{eq:6.25}
\end{eqnarray}
The limit of this amplitude turns out to be
\begin {eqnarray}
Z_{6'}^{(2)}(\theta ^k) &\to& -(1-1)\frac{1}{2}4t
(-2\sin \pi kv_3)\frac{(1+q_{(1/2t)}^{-kv_3})}{(1-q_{(1/2t)}^{-kv_3})}
\nonumber\\
&\times& \prod _{n=1}^{\infty}
\frac{(1-q_{(1/2t)}^{n-kv_3-\frac{1}{2}})(1-q_{(1/2t)}^{n+kv_3-\frac{1}{2}})(1+q_{(1/2t)}^{n-kv_3})(1+q_{(1/2t)}^{n+kv_3})}
{(1+q_{(1/2t)}^{n-kv_3-\frac{1}{2}})(1+q_{(1/2t)}^{n+kv_3-\frac{1}{2}})(1-q_{(1/2t)}^{n-kv_3})(1-q_{(1/2t)}^{n+kv_3})}\nonumber \\
&\times& \prod_{i=1,2}(2\cos \pi kv_i)\frac{(1-q_{(1/2t)}^{-kv_i})}{(1+q_{(1/2t)}^{-kv_i})}
\nonumber\\
&\times& \prod _{n=1}^{\infty}
\frac{(1+q_{(1/2t)}^{n-kv_i-\frac{1}{2}})(1+q_{(1/2t)}^{n+kv_i-\frac{1}{2}})(1-q_{(1/2t)}^{n-kv_i})(1-q_{(1/2t)}^{n+kv_i})}
{(1-q_{(1/2t)}^{n-kv_i-\frac{1}{2}})(1-q_{(1/2t)}^{n+kv_i-\frac{1}{2}})(1+q_{(1/2t)}^{n-kv_i})(1+q_{(1/2t)}^{n+kv_i})}
\label{eq:6.26}
\end{eqnarray}
When $kv_3 =$ integer, the part which contains $kv_3$ should be replaced with $(4t)(-1)^{kv_3}$. Furthermore, when $2kv_i = 2n+1, n=0,1,\cdots,$ the part which contains $kv_i, i=1,2$ should be replaced with $(4t)(-1)^{n+1}$.

Next we compute the contribution of ${\cal L}_3$ to $Z_p(\theta^k)$.
M\"{o}bius 6-states have NN boundary conditions for $\mu=4,6,8$ and DD boundary conditions for $\mu=5,7,9$.
When $kv_i=$integer, $|n,m\rangle$ transforms as Eq.(\ref{eq:6.9}) and $Z^{(3)}_6(\theta^k)$ reads
\begin {eqnarray}
Z_6^{(3)}(\theta ^k) &=& \prod_i \frac{1}{2t}\frac{L_{e_i}}{L_{o_i}}
\sum_s e^{-\pi \rho s^2/2t} \sum_s e^{-\pi s^2/2t\rho}
\nonumber\\
&\to& \prod_i \frac{1}{2t}\frac{L_{e_i}}{L_{o_i}}
\label{eq:6.27}
\end{eqnarray} 
When $kv_i=$half integer, $|n,m\rangle$ transforms as Eq.(\ref{eq:6.11}).
But M\"{o}bius 6-states have no winding $(m_{\mu}=0)$ for $\mu=4,6,8$ and no momentum $(n_{\mu}=0)$ for $\mu=5,7,9$.
So they have no contribution to $Z^{(3)}_6(\theta^k)$ .

M\"{o}bius $6'$-states have NN boundary conditions for $\mu=5,7,8$ and DD boundary conditions for $\mu=4,6,9$. When $kv_3 =$ integer, a sum over quantized momenta in $\mu = 8$ and a sum over windings in $\mu = 9$ contribute to the amplitude.
Thus, for $kv_3=$integer, we obtain
\begin {eqnarray}
Z_{6'}^{(3)}(\theta^k) &=& \frac{1}{2t}\frac{L_8}{L_9}
\sum_s e^{-\pi \rho s^2/2t} \sum_s e^{-\pi s^2/2t\rho}
\nonumber\\
&\to& \frac{1}{2t}\frac{L_8}{L_9}
\label{eq:6.28}
\end{eqnarray} 

When $kv_i=$ half integer for $i=1,2$, a sum of quantized momenta contributes in $\mu = 5, 7$, while a sum of windings contributes in $\mu = 4,6$.
So, if $kv_i =$ half integer for $i=1,2$, we have
\begin {eqnarray}
Z_{6'}^{(3)}(\theta^k) &=& \prod_i \frac{1}{2t}\frac{L_{o_i}}{L_{e_i}}
\sum_s e^{-\pi \rho s^2/2t} \sum_s e^{-\pi s^2/2t\rho}
\nonumber\\
&\to& \prod_i \frac{1}{2t}\frac{L_{o_i}}{L_{e_i}}
\label{eq:6.29}
\end{eqnarray} 
where $L_{o_i}=(L_5,L_7)$ and $L_{e_i}=(L_4,L_6)$.
\newpage

%

\begin{thebibliography}{99}
%
\bibitem{pol}
See, e.g., J.~Polchinski, {\it String Theory, Vol.II}, (Cambridge U.P., 1998).
%
\bibitem{pw}
J.~Polchinski and E.~Witten, \NPB {460}{96}{525}.
%
\bibitem{ginsparg}
P.~Ginsparg, \PRD {35} {86} {648}.
%
\bibitem{D=6_I}
G.~Pradisi and A.~Sagnotti, \PLB {216} {89} {59}; M.~Bianchi and A.~Sagnotti, \PLB {247} {90} {517}; \NPB {361} {91} {519}; E.G.~Gimon and C.V.~Johnson, \NPB {477} {96} {715}; A.~Dabholkar and J.~Park, \NPB {477} {96} {701}; C.~Angelantonj, M.~Bianchi, G.~Pradisi, A.~Sagnotti and Ya.S.~Stanev, \PLB {387} {96} {743}.
%
\bibitem{gp}
E.G.~Gimon and J.~Polchinski, \PRD {54} {96} {1667}.
%
\bibitem{D=6_IIB}
A.~Dabholkar and J.~Park, \NPB {472} {96} {207}; \PLB {394} {97} {302}; M.~Berkooz, R.G.~Leigh, J.~Polchinski, J.H.~Schwarz, N.~Seiberg and E.~Witten, \NPB {475} {96} {115}; J.~Polchinski, \PRD {55} {97} {6423}, E.G.~Gimon and C.V.~Johnson, \NPB {479} {96} {285}; Z.~Kakushadze, G.~Shiu and S.-H.~Henry Tye, \PRD {58} {98} {086001}.
%
\bibitem{D=4}
C.~Angelantonj, M.~Bianchi, G.~Pradisi, A.~Sabnotti and Ya.S.~Stanev, \PLB {385} {96} {96}; M.~Berkooz, R.G.~Leigh, \NPB {483} {97} {187};
G.~Zwart, "Four-dimensional $N=1$ $Z_N \times Z_M$ orientifolds", hep-th/9708040; 
Z.~Kakushadze, and G.~Shiu, \PRD {56} {97} {3686}; \NPB {520} {98} {75}; Z.~Kakushadze, \NPB {512} {98} {221}; \PLB {434} {98} {269}.
%
\bibitem{ald}
G.~Aldazabal, A.~Font, L.E.~Ibanez and G.Violero, "$D=4, N=1$, type IIB orientifolds", hep-th/9804026.
%
\bibitem{kak}
Z.~Kakushadze, "On four dimensional $N=1$ type I compactifications", hep-th/9806008.
%
\bibitem{dhvw}
L.~Dixon, J.A.~Harvey, C.~Vafa and E.~Witten, \NPB {274} {86} {285}.
%
\bibitem{imnq}
L.E.~Ib\'a\~nez, J.~Mas, H.-P.~Nilles and F.~Quevedo, \NPB {301} {88} {157}.
%
\bibitem{kst}
Z.~Kakushadze, G.~Shiu and S.-H.~Henry Tye, "Type IIB orientifolds, F-theory, type I strings on orbifolds and type I-heterotic duality", hep-th/9804092.
%
\bibitem{cp}
Y.~Cai and J.~Polchinski, \NPB {296}{88}{91}.
%
\bibitem{pol2}
J.~Polchinski, \PRL {75}{95}{4724}.
%
\bibitem{mum}
D.~Mumford, {\it The Tata Lectures on Theta I, Birkh\"auser}, 1983.
%
\bibitem{blum}
J.~Blum, "F theory orientifolds, M theory orientifolds, and twisted strings", hep-th/9608053.
%
\end{thebibliography}
\end{document}